\documentclass[journal]{IEEEtran}

\usepackage{amsfonts}
\usepackage{amsmath}
\usepackage{amssymb}
\usepackage{graphicx}
\usepackage{mathtools}
\usepackage{adjustbox}
\usepackage{hyperref}
\hypersetup{colorlinks,allcolors=black}
\usepackage{cite} 

\usepackage{amsthm} 

\usepackage{soul}  

\usepackage[normalem]{ulem}
\newtheorem{theorem}{Theorem}

\begin{document}
\title{Application of Guessing to Sequential Decoding of Polarization-Adjusted Convolutional (PAC) Codes}

\author{Mohsen Moradi\textsuperscript{\href{https://orcid.org/0000-0001-7026-0682}{\includegraphics[scale=0.06]{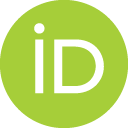}}},~\IEEEmembership{Member,~IEEE}
\thanks{
This paper is based on \cite[Ch.~4]{moradi2022performance}.
The author is with the Department of Electrical-Electronics Engineering, Bilkent University, Ankara TR-06800, Turkey (e-mail: moradi@ee.bilkent.edu.tr).}
}

\maketitle
\begin{abstract}
Despite the extreme error-correction performance, the amount of computation of sequential decoding of the polarization-adjusted convolutional (PAC) codes is random.
In sequential decoding of convolutional codes, the computational cutoff rate denotes the region between rates whose average computational complexity of decoding is finite and those which is infinite.
In this paper, by benefiting from the polarization and guessing techniques, we prove that the computational cutoff rate in sequential decoding of pre-transformed polar codes polarizes. 
The polarization of the computational cutoff rate affects the criteria for the rate-profile construction of the pre-transformed polar codes.
We propose a technique for taming the Reed-Muller (RM) rate-profile construction, and the performance results demonstrate that the error-correction performance of the PAC codes can achieve the theoretical bounds using the tamed-RM rate-profile construction and requires a significantly lower computational complexity than the RM rate-profile construction.
\end{abstract}


\section{Introduction}
\IEEEPARstart{C}{onnecting} polar and convolutional coding, polarization-adjusted convolutional (PAC) codes are a family of linear codes \cite{arikan2019sequential}. 
The encoder of PAC codes can be regarded as a tree code, allowing a sequential decoding algorithm to be used to decode the codewords. 

The sequential decoding of PAC codes has a variable computational complexity, and similar to the sequential decoding of conventional convolutional codes (CCs), it is susceptible to the cutoff rate phenomena \cite{moradi2020performance, moradi2021sequential}. 
In addition, the utilization of polar codes in the PAC codes brings about a distinction in the computational complexity analysis of decoding in comparison to the sequential decoding of CCs. 
The main objective of this paper is to provide a lower bound on the computational complexity of sequential decoding of PAC codes utilizing the guessing function \cite{arikan1996inequality}. 
Ar{\i}kan in \cite{arikan1996inequality} provided a tight lower bound on the average computation required for sequential decoding of conventional CCs by employing the relationship between the computational complexity of the sequential decoding and the guessing function. 
We address the computational complexity of sequential decoding of PAC codes using this lower bound and channel polarization approach.

Sequential decoding is a tree search algorithm introduced by Wozencraft \cite{wozencraft1957sequential} that performs the decoding by attempting to guess its path through an expanding tree of the most probable transmitted sequences.
The computational complexity would be reduced in this manner. 
In the case of sequential decoding of CCs, this generally comes at the cost of communicating at rates strictly below capacity. 
The computational cutoff rate denotes the region between rates whose average computational complexity of the sequential decoding is finite and those which is infinite.
This paper, benefiting from the polarized channels, proves that the computational cutoff rate in sequential decoding of PAC codes polarizes. 

The Fano \cite{fano1963heuristic} and the stack \cite{zigangirov1966some, jelinek1969fast} algorithms are both very well-known examples of sequential decoding algorithms. 
Fano algorithm may make many visits to the nodes of the decoding tree, while the stack algorithm visits each node of the decoding tree no more than once but requires a larger amount of storage space.
Since both algorithms ultimately choose the same pathways on the decoding tree, the Fano and stack methods visit the same set of nodes. 
In the simulations presented in this study, we use the Fano algorithm.

The rate profile and convolutional encoder used in the construction of a PAC code significantly impact its performance. 
In \cite{moradi2022tree}, to propose and analyze the metric function for the list decoding of PAC codes, the difficulty of the CC in the PAC code analysis is avoided by assuming that the output of the CC for the information bits is random.
Although a good convolutional encoder can significantly improve the error-correction performance of PAC codes, our simulation findings demonstrate that the convolutional encoder has almost no effect on the complexity performance. 
This is analogous to CCs in which, although the complexity of the Viterbi decoding grows exponentially with the code constraint length, the complexity of the sequential decoding is invariant to the code constraint length \cite{viterbi2013principles}. 
In this study, the rate-profile construction of PAC codes is examined. 

As the block length $N$ approaches infinity, our findings demonstrate that the rate profile of the PAC codes should fall below the cutoff rate profile. 
This implies that, in order to have a tractable sequential decoder, the PAC codes rate profile should be consistent with the polar rate profile. 
On the other hand, for $N=128$, it is shown that sequential decoding of the PAC code designed with the Reed-Muller (RM) rate profile can meet the theoretical bounds with low average computational complexity \cite{moradi2021sequential}.
Our simulation findings indicate that raising the block length to even $N=256$ may significantly increase the average computational complexity of sequential decoding. 
We propose an approach based on the polarization of the computational cutoff rate to tame the RM rate profile of the PAC codes.
Recently, much research has been done in order to suggest an algorithm for the rate-profile construction of the PAC codes.

Based on the cutoff rate polarization, the proposed method in \cite{moradi2021monte} attempts to enhance the error-correction performance of the PAC codes while ensuring a low average sequential decoding complexity for signal-to-noise ratio (SNR) values above a target SNR value.
In \cite{seyedmasoumian2022approximate}, the PAC$(64, 32)$ coding rate profile is designed using a discrete optimization technique based on simulated annealing, and the results indicate that PAC codes with this proposed rate profile have a high error-correction performance. 
It would be interesting to adapt this method to larger block lengths. 
In \cite{mishra2022modified}, an approach for reinforcement learning-based rate-profile construction is presented.
This method employs a collection of reward and update mechanisms that allow the reinforcement learning agent to determine the rate profile. 
In \cite{tonnellier2021systematic}, it is demonstrated that a PAC$(256, 128)$ code may also meet the theoretical limits by using a genetic approach to obtain the code rate profile. 
As this coding design targets only the error-correction performance of the code, the computational complexity can be extremely high.
A randomized construction of polar subcodes is presented in \cite{trifonov2017randomized}, with the objective of minimizing the complexity of low-weight codewords in the resulting codes and boosting performance under list or sequential decoding. 
The results demonstrate that, with this construction, stack decoding is less complex than polar codes with CRC. 
The SC-Flip \cite{chandesris2018dynamic} decoding can likewise be seen as an online rate profile construction approach that adjusts the incorrect bits during decoding.  
Similar to sequential decoding, in SC-Flip decoding, the complexity is reduced at high SNR levels as there are fewer erroneous bit locations.

In \cite{moradi2021monte}, to construct a PAC$(N, K)$ code, more than $K$ reliable subchannels are picked as the indices for the information bits, and the more erroneous ones are frozen one-by-one during decoding; but, this approach is yet empirical. 
In \cite{liu2022weighted}, the algorithm begins by picking subchannels with high weights and then updates the subchannels with lower weights depending on the subchannels' reliability via repeated encoding.
As the initialization does not dependent on the reliability of the subchannels, the rate profile construction results in extremely high decoding complexity.

In this paper, we use boldface letters to denote vectors and matrices. 
All operations are over a binary field $\mathbb{F}_2$. 
We use $\mathbf{u}^i$ to denote subvector $(u_1, \cdots, u_i)$ and $\mathbf{u}_i^j$ to denote subvector $(u_i, \cdots, u_j)$.

The remainder of this paper is organized as follows.
Section \ref{sec: background} briefly reviews polar codes and channel polarization.
Section \ref{sec: RM} discusses the RM codes.
Section \ref{sec: scheme} gives an overview of the parameters and blocks of the PAC codes and metric function used in this paper.
The sequential decoding of PAC codes is detailed in Section {\ref{sec: seq dec}}.
In Section \ref{sec: guessing}, the polarization of the computational complexity is proved. 
Section \ref{sec: simulation} provides simulation results.
Finally, Section \ref{sec: conclusion} concludes this paper.


\section{Background on Polar codes} \label{sec: background}
This section briefly covers polar code encoding and decoding.
Let $W: \mathcal{X} \longrightarrow \mathcal{Y}$ denotes a binary input discrete memoryless channel (B-DMC) with arbitrary output alphabet $\mathcal{Y}$. 
The channel transition probability is defined by $W(y|x)$, where $x \in \mathcal{X}$ and $y \in \mathcal{Y}$.
The generator matrix of polar codes can be obtained from the rows of $F_N \triangleq \mathbf{F}^{\otimes n}$, which is the $n$th Kronecker power of $\mathbf{F} = \begin{bsmallmatrix} 1 & 0\\ 1 & 1 \end{bsmallmatrix}$ with $n = \log_2 N$. 
Determining this submatrix corresponds to the selection of the most reliable subchannels $W_N^{(i)}: \mathcal{X} \longrightarrow \mathcal{Y} \times \mathcal{X}^{i-1}$ as explained in \cite{arikan2009channel}.

For an $(N, K, \mathcal{A})$ polar code with $N = 2^n$, the $K$ information vector $\mathbf{d}^K$ of length $K$ first can be inserted into the vector $\mathbf{u}^N$ as $\mathbf{u}_{\mathcal{A}} = \mathbf{d}^K $ and $\mathbf{u}_{\mathcal{A}^{c}} = \mathbf{0}$.
The complementary set $\mathcal{A}^c$ denotes the frozen bit set, and the frozen bits $\mathbf{u}_{\mathcal{A}^c}$ can be assigned to all zeros for the symmetric channels. 
Then, encoding is done as $\mathbf{x}^N = \mathbf{u}^N \mathbf{F}^{\otimes n}$.

An important parameter of channel $W$ is the Bhattacharyya parameter which is defined as
\begin{equation}
    Z(W) \triangleq \sum_{y \in \mathcal{Y}} \sqrt{W(y|0)W(y|1)} .
\end{equation}

Selecting the most reliable subchannels and determining the information set $\mathcal{A}$ is to calculate the bit-channel Bhattacharyya values $Z(W_N^{(i)})$ and choose the channels with the least bit-channel Bhattacharyya values.

\section{Background on RM codes} \label{sec: RM}
Reed-Muller (RM) codes are a family of linear block codes having a simple construction and rich structural properties \cite{muller1954application, reed1953class}.
For all integers $m$ and $r$ ($0 \leq r \leq m$), there exists an $r$th-order RM code, denoted as RM$(r,m)$, with a code length of $N=2^r$ and the dimension of 
$K(r,m) = {m \choose 0}+{m \choose 1}+\cdots {m \choose r}$,
where ${m \choose i}$ is the binomial coefficient.
An RM$(r,m)$ code is constructed by selecting all $K(r,m)$ row indices of the matrix $\mathbf{F}^{\otimes r}$ with the Hamming weights more than or equal to $d_{\text{min}} = 2^{m-r}$, where $d_{\text{min}}$ is the code minimum distance.
Note that one difference between polar and RM codes is how the row indices are selected. 
Also, in an RM code, the code dimension $K$ can take on $m+1$ distinct values, whereas in polar codes $1\leq K \leq N$. 
We occasionally use RM$(N, K)$ notation rather than the more standard RM$(r,m)$ notation.

Despite being an old family of error-correcting codes that have been theoretically investigated very well, RM codes have attracted a growing number of scholars in recent years \cite{kamenev2021sequential, thangaraj2020efficient, dumer2006soft, hashemi2018decoding, geiselhart2021automorphism, santi2018decoding, ye2020recursive, lian2020decoding, abbe2020reed, li2019pre, li2021performance}.
\begin{figure}[t] 
\centering
	\includegraphics [width = \columnwidth]{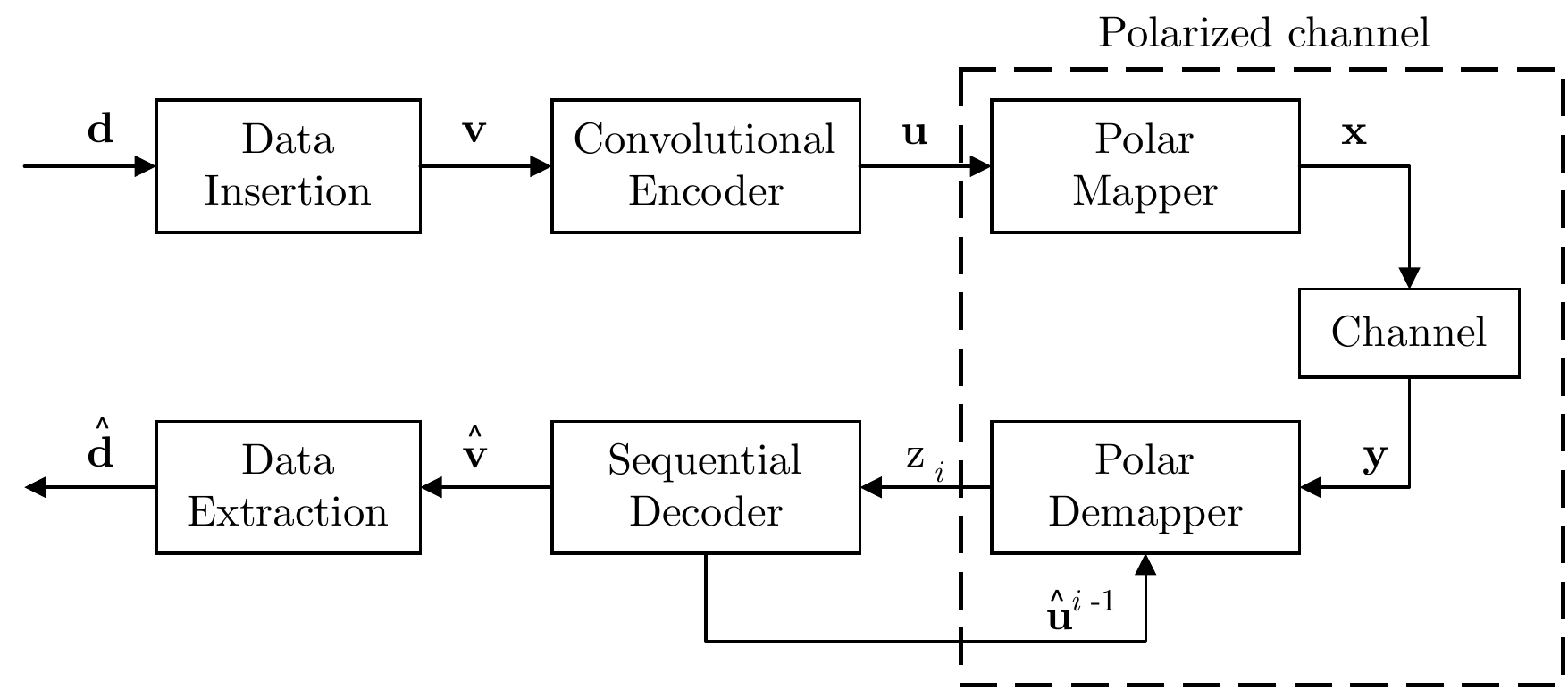}
	\caption{Flowchart of PAC coding scheme.} 
	\label{fig: flowchart}
\end{figure}

\section{PAC Coding Scheme} \label{sec: scheme}
Fig. \ref{fig: flowchart} shows a block diagram of the PAC coding scheme.
For an $(N,K,\mathcal{A}, \mathbf{T})$ PAC code, the parameters $N$ and $K$ are specified same as polar code parameters, $\mathcal{A}$ is the PAC code rate profile, and $\mathbf{T}$ is an upper-triangular Toeplitz matrix constructed with a connection polynomial $\mathbf{g}(x) = g_mx^{m} + \cdots + g_1x + g_0$, with $g_0 = g_m = 1$ represented as

\begin{equation*}
\mathbf{T} = 
\begin{bmatrix}
 g_0    & g_1    &  g_2   & \cdots & g_m    & 0      & \cdots & 0      \\
 0      & g_0    & g_1    & g_2    & \cdots & g_m    &        & \vdots \\
 0      & 0      & g_0    & g_1    & \ddots & \cdots & g_m    & \vdots \\
 \vdots & 0      & \ddots & \ddots & \ddots & \ddots &        & \vdots \\
 \vdots &        & \ddots & \ddots & \ddots & \ddots & \ddots & \vdots \\
 \vdots &        &        & \ddots & 0      & g_0    & g_1    & g_2    \\
 \vdots &        &        &        & 0      & 0      & g_0    & g_1    \\
 \vdots & \cdots & \cdots & \cdots & \cdots & 0      & 0      & g_0    
\end{bmatrix}.
\end{equation*}

The vector $\mathbf{d}^K$ is the source word generated uniformly at random over all possible source data of length $K$ in a binary field $\mathbb{F}_2$.
The data insertion (rate profile) block maps these $K$ bits into a data carrier vector $\mathbf{v}^N$ in accordance with the data set $\mathcal{A}$, thus inducing a code rate of $R=K/N$.
After $\mathbf{v}^N$ is obtained by $\mathbf{v}_\mathcal{A} = \mathbf{d}^K$ and $\mathbf{v}_{\mathcal{A}^c} = \mathbf{0}$, it is encoded as $\mathbf{u}^N = \mathbf{v}^N\mathbf{T}$ by the convolutional encoder.
This entails a constraint on each component $u_j$ of the vector $\mathbf{u}^N$ that is dependent on the at most $m$ bits that come before it.
These bits, $u_j$, can be regarded as the dynamic frozen (in the dynamic frozen bits, each frozen bit corresponds to the linear combination of bits with lower indices) or parity check frozen symbols \cite{trifonov2013polar, wang2016parity}. 
The vector $\mathbf{u}^N$ is finally encoded using the polar mapper (polar transformation) as $\mathbf{x}^N = \mathbf{u}^N \mathbf{F}^{\otimes n}$.
The sequential decoder receives polarized channel output $z_i$ (LLR value corresponding to the $\hat{u}_i$ \cite{moradi2021sequential}) and outputs an estimate of $\hat{v}_i$. 
At the end of decoding, from $\hat{\mathbf{v}}^N$ the $K$-bits data can be extracted according to $\mathcal{A}$.
Employing sequential decoding for polarized channels is described in the next section in detail.

Numerous research has examined the sequential decoding of polar-like codes using various heuristic metric functions.
The metric function in \cite{moradi2020performance} uses fixed bias values for different coding rates. 
This fixed bias value can be calculated by adding the bit-channel cutoff rate values and dividing by $N$. 
As the fixed bias value disregards polarized channels, it may result in a high level of computational complexity. 
An heuristic path metric function for the SC-Fano decoding for polar codes with a fixed bias value is also proposed in \cite{jeong2019sc}. 
The study \cite{niu2012stack} presents a path metric function for sequential decoding of polar codes that updates the metric only for the branching levels. 
As an extension of \cite{niu2012stack}, \cite{wang2016parity} additionally updates the path metric for frozen bits. 
In \cite{miloslavskaya2014sequential} and \cite{trifonov2018score}, the proposed metric functions utilize codeword probability estimations of the most probable codeword to determine the continuation of a code tree path. 
Based on an investigation of the bit-metric function of the list decoding, \cite{moradi2022tree} presents a pruning strategy such that, similar to sequential decoding, the average complexity of the list decoding converges to one per bit decoded.

\section{Sequential decoding} \label{sec: seq dec}
A sequential decoding algorithm searches the code tree of the PAC code for the correct path that corresponds to the transmitted data. 
The key principle of sequential decoding is that only the most promising paths should be considered during the decoding process. 
If it seems that a path leading to a node is not reliable, the decoder may reject all paths emanating from that node without suffering a large performance loss compared to the maximum-likelihood (ML) decoding.
The path metric function directs a decoder to examine the most likely path.

Fano \cite{fano1963heuristic} and stack \cite{zigangirov1966some, jelinek1969fast} algorithms are two well-known sequential decoding algorithms. 
The stack algorithm creates a stack of already traversed pathways of varied lengths, ordered by their metric values in decreasing order. 
A drawback with this algorithm is that it is always possible for the stack to get too large before decoding a given frame, and any application of stack decoding must have a stack size limit, resulting in a loss of performance.
Another concern is the reordering of the stack after each decoding step, which is a function of the number of existing paths in the stack.
As the number of elements in the stack increases, this may seriously affect the decoding delay. 
Due to channel polarization, the metric value of the correct path may be well distinguishable from the wrong paths in PAC codes.
At a high SNR, this can decrease the number of elements in the stack to almost one, hence solving the sorting issue \cite{moradi2022tree}.

The Fano algorithm is widely recognized as the most practical sequential decoding algorithm, and it examines a single path at a time, eliminating the need to store anything other than a single path and its metric value.
In essence, the algorithm explores a path so long as its metric value increases.
When the metric value begins to decrease significantly, the algorithm returns to earlier nodes on previously traveled paths and seeks other paths that stem from them. 
The algorithm makes use of a series of comparison thresholds denoted by $T$ that are spaced apart by $\Delta$ values.
When the metric value increases enough during forward searching, the threshold is raised by $\Delta$ and lowered by $\Delta$ during backward searching. 
This is conducted in such a way that no node is ever searched forward twice with the same threshold value; the threshold should always be less than the previous value.

The search complexity of a sequential decoding algorithm is a random variable that is mostly dependent on the level of the noise. 
We define this complexity using a random variable, $\Theta$, that counts the number of nodes the decoder accesses throughout a decoding session.
We are interested in the expectation of this random variable per decoded bit (i.e., $\mathbb{E}[\Theta]/N$), often known as the average number of visits (ANV) \cite{moradi2020performance}. 
Note that the ANV in \cite{moradi2020performance} is per codeword.
We utilize per decoded bit since it has the same unit as the list size in the SC list decoding.
The stack algorithm may only visit each node of the code tree once, but due to the backtracking characteristic of the Fano algorithm, it may visit some nodes many times, and $\Theta$ counts every one of these visits. 

\subsection{Decoding of PAC Codes}
Two building blocks comprise the decoding of a PAC code: the polar demapper and the sequential decoder. 
In order to decode a PAC code, we explain how to adopt the Fano algorithm as a sequential decoding technique.

Assume that the Fano decoder is moving towards the $i$th node in a forward manner. 
As with the SC decoder, the polar demapper receives the channel output $\mathbf{y}$ and calculates the soft output 
\begin{equation}
    z_i \triangleq \log_2 \left( \dfrac{P(\mathbf{y}^N,\hat{\mathbf{u}}^{i-1}|u_i = 0 )}{P(\mathbf{y}^N,\hat{\mathbf{u}}^{i-1}|u_i = 1 )}
    \right),
\end{equation}
using the hard decisions $\hat{\mathbf{u}}^{i-1}$ vector supplied by the sequential decoder. 
Note that, unlike the SC decoder, the polar demapper does not make a hard decision but instead provides the sequential decoder with the soft $z_i$ values. 

The bit metric for the $i$th branch of the Fano decoder is given by
\begin{equation}\label{eq: bitmetricformula}
\begin{split}
    & \gamma(u_{i} ;\mathbf{y}^N,\mathbf{u}^{i-1}) = 1 - \log_2 \left(
    1 + 2^{- z_i \mathord{\cdot}(-1)^{u_i}}
    \right) - b_i,
\end{split}
\end{equation}
where $(\mathbf{y}^N,\mathbf{u}^{i-1})$ is the output of the $i$th polarized channel $W_N^{(i)}$, $u_i$ is the branch of the tree at the $i$th level, and $b_i$ is the bias value of the $i$th bit \cite{moradi2021sequential}. 
The bit-channel bias value $b_i$ is a design parameter, and in the simulations of this paper, we use the bit-channel cutoff rates \cite{moradi2021sequential} for $b_i$, which for the path metric results in an average positive drift for the correct path and a negative drift for the wrong directions.

Using this metric function, Fano decoding obtains $\hat{v}_i$.
In addition, the Fano decoder obtains $\hat{\mathbf{u}}^i$ from $\hat{\mathbf{v}}^i$ (through an encoder replica) and delivers $\hat{\mathbf{u}}^i$ to the polar demapper.
Then, using $\hat{\mathbf{u}}^i$, the polar demapper calculates $z_{i+1}$, and the decoding process continues until $\hat{v}_N$ is obtained or a specified stopping rule ends the decoding procedure.

As was noted before, the Fano decoder has to go backwards if the tentative path metrics of both children are lower than the running threshold $T$ and the path metric of the node that came before it is higher than $T$.
Consider that the Fano decoder is now placed at the $i$th node and intends to backtrack to the preceding $(i-1)$th node by feeding $\hat{\mathbf{u}}^{i-1}$ to the polar demapper.
To prevent the polar demapper from starting the demapping operation again from scratch when calculating $z_{i-1}$, it is required to preserve all intermediate LLR values; to do so, the polar demapper would try to move backward from the common ancestor of the $i$th and $(i-1)$th nodes. 
In a similar manner, for the polar demapper to go backward from the $i$th node to the $j$th node when $ i>j$, it is sufficient for the polar demapper to begin from the common origin of the $i$th and $j$th leaf nodes of the polar demapper tree. 
Consequently, the explained polar demapper retains all intermediate LLRs and has a memory capacity of $N\log_2{N}$.
Polar demapper has a trade-off between delay and memory usage, and storing only $N-1$ intermediate LLR values incurs a considerable latency increase owing to the backtracking aspect of the Fano algorithm \cite{moradi2021sequential}.

\section{Guessing and sequential decoding} \label{sec: guessing}
 
In information theory, guessing traces its origins to Massey's work \cite{massey1994guessing}.
Massey proved that by guessing the value of a random variable $X$ in decreasing order of the probabilities $(p_1 \geq p_2 \geq \cdots)$, the number of guesses $G(X)$ would have the smallest average, where $p_1$ is the probability of the most likely symbol in the space of the random variable $X$, $p_2$ is for the second most likely symbol and so on.
In this manner, for $H(X)\geq 2$
\begin{equation}
    \mathbb{E}[G(X)] = \sum_{i=1} i.p_i \geq \frac{1}{4}2^{H(X)}+1,
\end{equation}
where $H(X)$ denotes the entropy function.

Ar{\i}kan \cite{arikan1996inequality} proved that for the random variable $X$ with a finite alphabet $\mathcal{X}$ of size $M$, by guessing the values in a decreasing order of the probabilities, the average number of successive guesses is upper and lower bounded as
\begin{equation}
    \left[ 
    \sum_{x\in \mathcal{X}}\sqrt{P_X(x)}
    \right]^2 \geq
    \mathbb{E}[G(X)] \geq \frac{\left[ 
    \sum_{x\in \mathcal{X}}\sqrt{P_X(x)}
    \right]^2}{1+\ln{M}},
\end{equation}
where $P_X(.)$ is the probability distribution of $X$. 

For a generalization that is useful in the channel coding problem \cite{arikan1996inequality}, consider a pair of discrete random variables $(X, Y)$ of the input and output of the channel where $X$ has probability distribution $P_X$ and takes one of the values in $\mathcal{X} = \{1, 2, \cdots, M \}$, and the channel output alphabet $\mathcal{Y}$ can be continuous.
For a given $Y$, the number of successive guesses needed to guess the correct input $X$, denoted by $G(X|Y)$, has a lower bound on its average as
\begin{equation}
    \mathbb{E}[G(X|Y)] \geq \frac{\sum_{y\in \mathcal{Y}}\left[ 
    \sum_{x\in \mathcal{X}}\sqrt{P_{X,Y}(x,y)}
    \right]^2}{1+\ln{M}}.
\end{equation}
where $P_{X,Y}(x,y)$ is the joint probability distribution of $(X, Y)$ \cite{arikan1996inequality}. 
Since the distribution $P_X$ is uniform and the size of $\mathcal{X}$ is equal to $M = e^{NR}$, the lower bound on the average of $G(X|Y)$ can be expressed as
\begin{equation}
    \mathbb{E}[G(X|Y)] \geq \frac{e^{NR - R_0(W)}}{1+NR}
    ,
\end{equation}
where $R_0(W)$ is the cutoff rate function for $(X,Y)$ and is defined as
\begin{equation}
    R_0(W) = R_0(X,Y) = -\log \sum_{y\in \mathcal{Y}}\left[ \sum_{x\in \mathcal{X}} P(x) \sqrt{P(y|x)} \right]^2 .
\end{equation}
Since, $\mathbb{E}[G(X|Y)]$ is similarly upper bounded by $e^{NR - R_0(W)}$, we consider $e^{NR - R_0(W)}$ as a tight lower bound and we use the notation
\begin{equation} \label{eq: guess_lb}
    \mathbb{E}[G(X|Y)] \gtrapprox e^{NR - R_0(W)}.
\end{equation}
The cutoff rate function is related to the Bhattacharyya parameter by
\begin{equation}
    R_0(W) = \log_2 \frac{2}{1+Z(W)},
\end{equation}
and consequently polarizing the Bhattacharyya parameter results in a polarized cutoff rate. 

To relate the number of guesses $G(X|Y)$ to sequential decoding of a PAC code, consider an arbitrary tree code of a PAC code and suppose that $\mathcal{X}$ is the set of all nodes at a fixed but arbitrary level $N$ of the tree, and $X$ is a random variable on $\mathcal{X}$ with a uniform distribution. 
We can think of $X$ as the node in $\mathcal{X}$, which lies on the transmitted path or equivalently as the channel input sequence of length $N$. 
The number of paths from the root of length $N$ is equal to the number of nodes at level $N$, and there is a one-to-one correspondence between them.
In this manner, the guessing function $G(.|.)$ is the sum of the number of nodes in $\mathcal{X}$ which are examined before, and the correct node $X = \mathbf{x}$ when $\mathbf{y}$ is received. 
No guess will be repeated in guessing the channel input, and whenever the correct channel input is guessed, the genie tells the decoder to stop.
Thus, the number of guesses $G(X|Y)$ is a lower bound to the PAC decoder's computation in decoding the first $N$ bits of the transmitted sequence.
Then, the lower bound to the average of guessing number $G(X|Y)$ serves as a lower bound to average computation in sequential decoding.

\begin{figure}[t] 
\centering
	\includegraphics [width = 0.7\columnwidth]{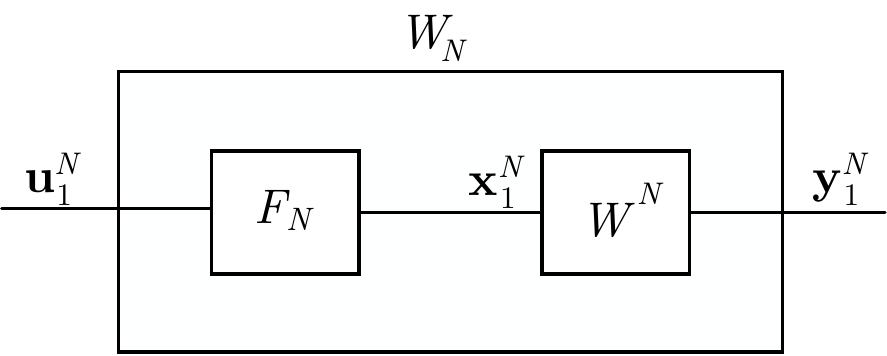}
	\caption{Polar code construction of length $N$.} 
	\label{fig: polarize_N1}
\end{figure}

As Fig.  \ref{fig: polarize_N1} illustrates, the combined channel $W_N$ that vector $\mathbf{u}^N$ sees is derived from a pre-processing on $N$ parallel channels seen by the vector $\mathbf{x}^N$.
Using $N$ copies of channel $W$, channel $W_N$ is obtained by the channel combining phase explained in \cite{arikan2009channel}.
Input-output pair of the channel $W_N$ is $(U^N; Y^N)$ and with Gallager's parallel channel theorem, the upper bound on the combined channel cutoff rate \cite[p.~149-150]{gallager1968information} we have
\begin{equation} \label{eq: Gal_parallel}
    R_0(W_N) = R_0(U^N; Y^N) \leq N R_0(W),
\end{equation}

\begin{figure}[h] 
\centering
	\includegraphics [width = .7\columnwidth]{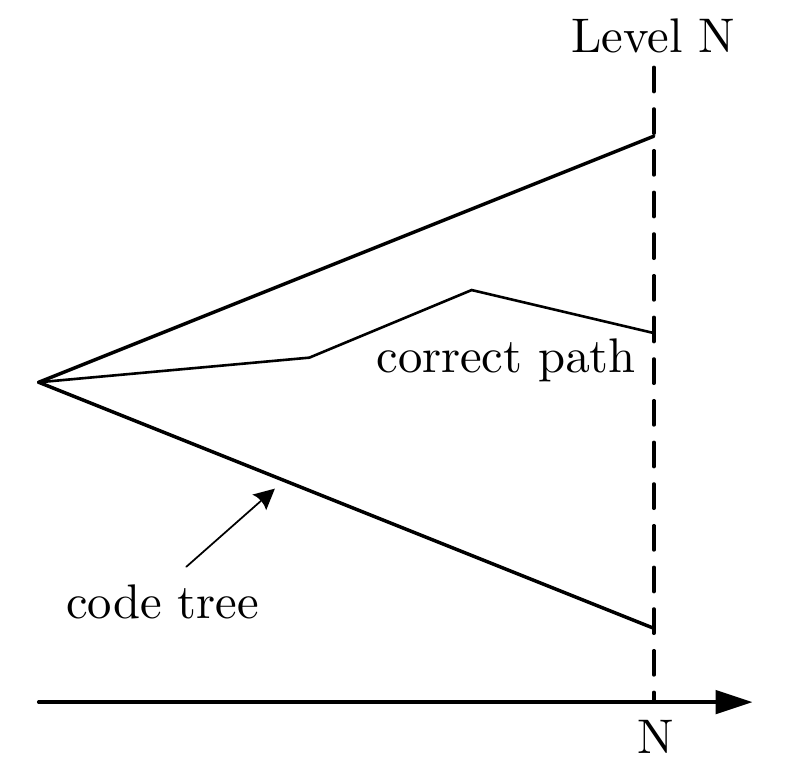}
	\caption{Decoding tree of PAC codes.}
	\label{fig: level_N1}
\end{figure}

Consider a $(N, K, \mathcal{A}, \mathbf{T})$ PAC code with the tree code shown in Fig.  \ref{fig: level_N1}.
By using \eqref{eq: guess_lb}, the average number of guesses $\mathbb{E}[G_{N}]$ has a lower bound
\begin{equation}
    \mathbb{E}[G_{N}] \gtrapprox e^{NR - R_0(U^N;Y^N)},
\end{equation}
and using \eqref{eq: Gal_parallel}, the lower bound on the average number of guesses becomes
\begin{equation} \label{eq: N_guess_LB}
    \mathbb{E}[G_{N}]  \gtrapprox e^{N(R-R_0(W))}.
\end{equation}

\begin{figure}[t] 
\centering
	\includegraphics [width = \columnwidth]{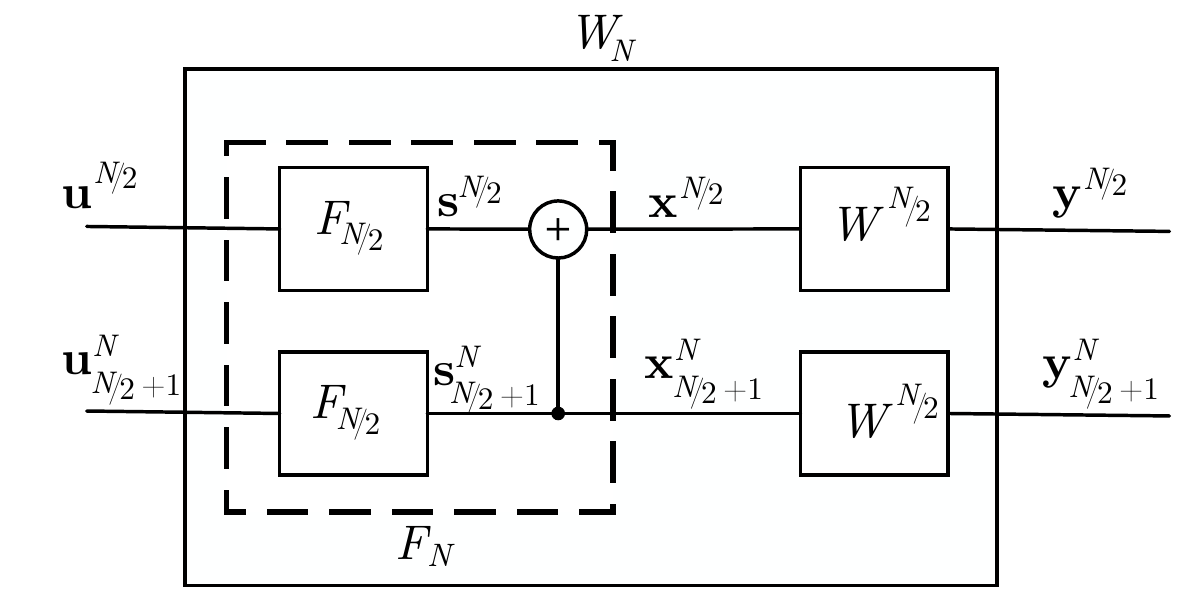}
	\caption{Recursive construction of polar code of length $N$.} 
	\label{fig: polarize_N2}
\end{figure}

The overall recursion of the polar mapper is illustrated in Fig.  \ref{fig: polarize_N2}.
We have $N$ parallel channels $W: X_i \rightarrow Y_i$, for  $1\leq i \leq N$. Suppose that $K^-$ is the number of information bits in $\mathbf{v}^{N/2}$ and $K^{+}$ is the number of information bits in $\mathbf{v}_{N/2 +1}^{N}$ s.t. $K = K^- + K^+$, and define
\begin{equation}
    R^- \triangleq \frac{K^-}{N/2}, ~~~ R^+ \triangleq \frac{K^+}{N/2}.
\end{equation}
Note that $R^- + R^+ = 2R$.
Similarly, let us denote the first and second halves' cutoff rates after one step of polarization by $R_0(W^-)$ and $R_0(W^+)$, respectively.
From the channel polarization theorem 
\begin{equation}
    R_0(W^-) + R_0(W^+) \geq 2R_0(W),
\end{equation}
which shows that after one step of polarization cutoff is boosted \cite{arikan2015origin}.
Polarization of the Gallager's function is also proved in \cite{alsan2014polarization}.
The main idea of boosting the cutoff rate is to build correlated synthesized channels of independent channels such that the sum of the cutoff rates of synthesized channels becomes greater than the independent channels.

\begin{figure}[h] 
\centering
	\includegraphics [width = 0.8\columnwidth]{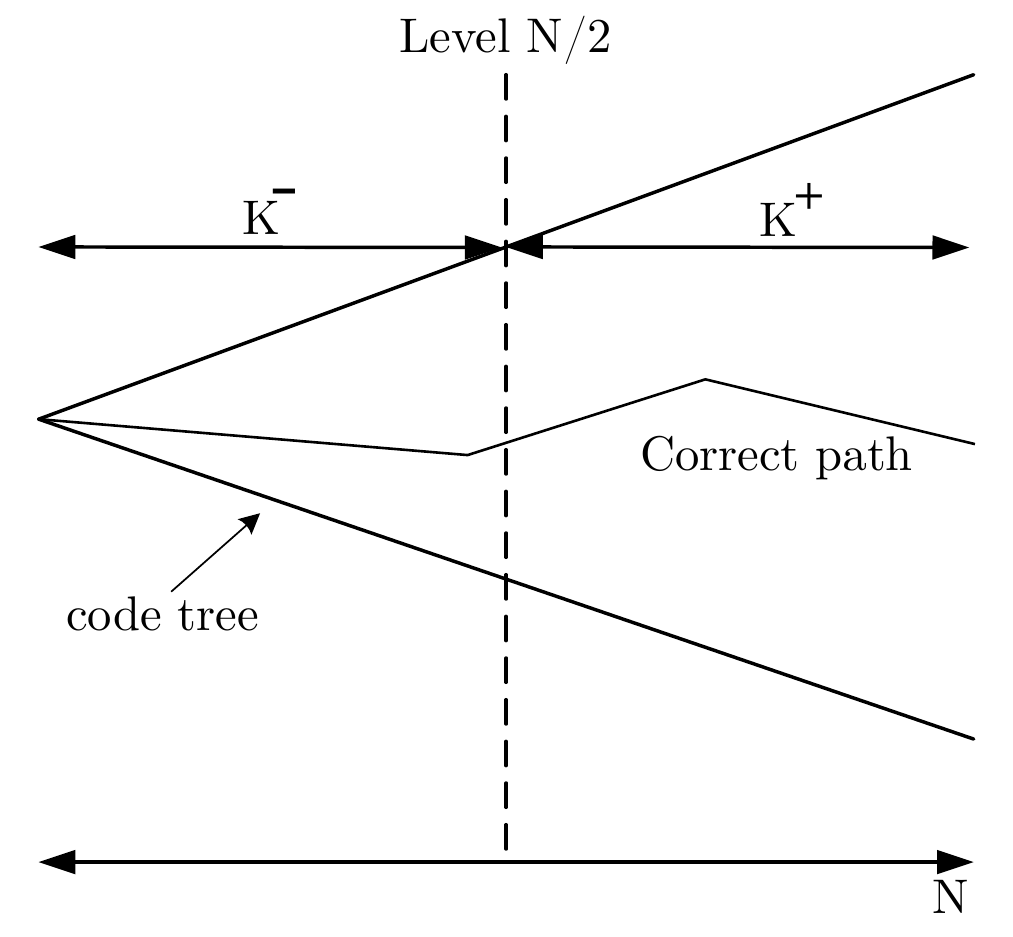}
	\caption{Decoding tree of PAC codes after one step polarization.} 
	\label{fig: level_N2}
\end{figure}

Suppose that the decoder in the tree code of Fig.  \ref{fig: level_N2} wants to reach the level $N/2$. 
We show the required average number of guesses by $\mathbb{E}[G_{N/2}]$. 
We also show the average number of guesses needed to decode the second half of the code as $\mathbb{E}[G_{N/2, N}]$ assuming a genie gives us the $\mathbf{u}^{N/2}$. 

\begin{theorem}
In sequential decoding of PAC codes, the computational cutoff rate polarizes, meaning that the lower bound on the average number of guesses for decoding the first and second halves of the codeword are exponential in $N/2$ as
\begin{equation}
\begin{split}
    &\mathbb{E}[G_{N/2}] \gtrapprox e^{\frac{N}{2}(R^{-}-R_0(W^-))},\\
    & \mathbb{E}[G_{N/2,N}] \gtrapprox e^{\frac{N}{2}(R^{+}-R_0(W^+))}.
\end{split}
\end{equation}
\end{theorem}

\begin{proof}
In one step polarization, we obtain $N/2$ parallel bad channels as
\begin{equation}
    W^-: S_i \rightarrow (Y_i,Y_{N/2 + i}),
\end{equation}
for $1\leq i \leq N/2$.
Suppose that the decoder in the tree code of Fig.  \ref{fig: level_N2} wants to reach the level $N/2$. 
The required average number of guesses has a lower bound as
\begin{equation}
    \mathbb{E}[G_{N/2}] \gtrapprox e^{(\frac{N}{2}R^{-}-R_0(U^{N/2};Y^N))}.
\end{equation}

Same as our first step, we have $N/2$ parallel copies of $W^-$, and a $F_{N/2}$ preprocessing is performed on the channel inputs $\mathbf{s}^{N/2}$ to obtain $\mathbf{u}^{N/2}$. 
As a result, by using the parallel channel theorem for the first half of the bit channels, we have
\begin{equation}
    R_0(U^{N/2};Y^{N}) \leq \frac{N}{2}R_0(W^-).
\end{equation}
Consequently, the average number of guesses required to decode the first half of the bits $\mathbb{E}[G_{N/2}]$ has the lower bound
\begin{equation}\label{eq: badN/2}
    \mathbb{E}[G_{N/2}] \gtrapprox e^{\frac{N}{2}(R^{-}-R_0(W^-))}.
\end{equation}

Moreover, in one step polarization, we also obtain $N/2$ parallel good channels  
\begin{equation}
    W^+: S_{N/2 +i} \rightarrow (Y_i,Y_{N/2 + i},S_i),
\end{equation}
for $1\leq i \leq N/2$.
In the same manner, if a genie provides the $\mathbf{u}^{N/2}$, the cutoff rate for the second half is obtained 
\begin{equation}
    R_0(U_{N/2 +1}^{N};Y^{N},U^{N/2}) \leq \frac{N}{2}R_0(W^+).
\end{equation}

With the genie-aided decoding assumption for the first half, the average number of guesses required to decode the second half has a lower bound 
\begin{equation}
    \mathbb{E}[G_{N/2,N}] \gtrapprox e^{(\frac{N}{2}R^{+}-R_0(U_{N/2 +1}^{N};Y^N,U^{N/2}))}.
\end{equation}
Therefore, the average number of guesses required to decode the second half of the bits $\mathbb{E}[G_{N/2,N}]$ has the lower bound
\begin{equation}\label{eq: goodN/2}
    \mathbb{E}[G_{N/2,N}] \gtrapprox e^{\frac{N}{2}(R^{+}-R_0(W^+))}.
\end{equation}
The lower bound on the number of guesses in \eqref{eq: N_guess_LB} is exponential in blocklength $N$, and after one step polarization \eqref{eq: badN/2} and \eqref{eq: goodN/2} are exponential in $N/2$ which is the gain in computational complexity of the PAC sequential decoder.
This proves that the computational cutoff rate polarizes.

\end{proof}

\begin{figure}[t] 
\centering
	\includegraphics [width = 0.9\columnwidth]{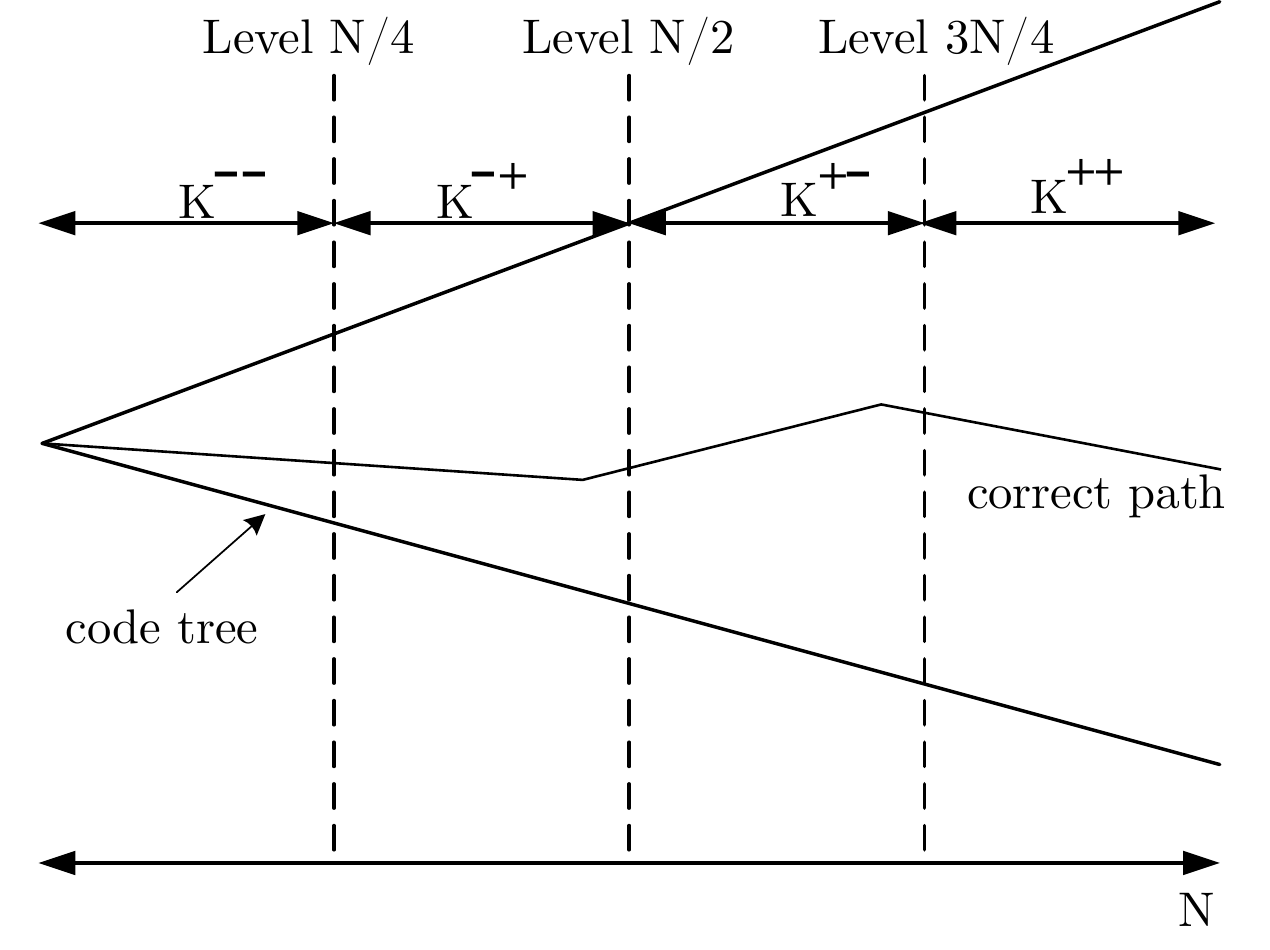}
	\caption{Decoding tree of PAC codes after two step polarization.} 
	\label{fig: level_N4}
\end{figure}

Fig.  \ref{fig: level_N4} extends the above operation recursively for the levels of size $N/4$. 
$K^{--}$ denotes the number of information bits in the first $N/4$ bits and $R^{--}$ is the corresponding rate. 
$K^{-+}$, $K^{+-}$, and $K^{++}$ are defined likewise with their corresponding rates. 
From channel polarization we have
\begin{equation}
    R_0(W^{--}) + R_0(W^{-+}) \geq 2R_0(W^{-}),
\end{equation}
and
\begin{equation}
    R_0(W^{+-}) + R_0(W^{++}) \geq 2R_0(W^{+}),
\end{equation}
which results in
\begin{equation}
    R_0(W^{--}) + R_0(W^{-+}) + R_0(W^{+-}) + R_0(W^{++}) \geq 4R_0(W).
\end{equation}

Following this procedure implies that the PAC code rate profile should be lower than the channel cutoff rate profile in order to have a minimal lower bound.

In \eqref{eq: Gal_parallel}, we have equality if the inputs to the polar mapper are independent.
To see this, assume $N=2$.
For $\mathbf{u} = (u_1, u_2)$ we have that 
\begin{equation}
    R_0(W_2) = 
     -\log \sum_{\mathbf{y}\in \mathcal{Y}\times \mathcal{Y}}\left[ \sum_{\mathbf{u}\in \mathcal{X}\times \mathcal{X}} P(\mathbf{u}) \sqrt{P(\mathbf{y}|\mathbf{u})} \right]^2,
\end{equation}
where $P(\mathbf{u})$ is a probability assignment on the input pairs.
If we restrict $P(\mathbf{u}) = P_1(u_1)P_2(u_2)$, where $P_1(.)$ and $P_2(.)$ are arbitrary input probability assignments on each parallel channel, then
\begin{equation}
\begin{split}
    R_0(W_2)& = -\log \sum_{\mathbf{y}\in \mathcal{Y}\times \mathcal{Y}} 
     \left(\sum_{u_1 \in\mathcal{X}}P_1(u_1)\sqrt{P(y_1|u_1)} \right)^2 \\
    &~~~~~~~~~~~~~~~~~~~ \times \left(\sum_{u_2 \in\mathcal{X}}P_2(u_2)\sqrt{P(y_2|u_2)} \right)^2\\
    & = -\log \sum_{y_1\in \mathcal{Y}}
    \left(\sum_{u_1 \in\mathcal{X}}P_1(u_1)\sqrt{P(y_1|u_1)} \right)^2\\
    &~~~ -\log \sum_{y_2\in \mathcal{Y}}
    \left(\sum_{u_2 \in\mathcal{X}}P_2(u_2)\sqrt{P(y_2|u_2)} \right)^2\\
    & = 2R_0(W).
\end{split}
\end{equation}
This is met by utilizing the standard ensemble of random codes for linear codes, which is detailed in \cite[p 206]{gallager1968information}.
This ensemble is designated by a fixed but arbitrary pair $(\mathbf{T},\mathbf{c})$ as $\mathbf{u} = \mathbf{v}\mathbf{T} + \mathbf{c}$, for the PAC codes.
$P(\mathbf{u}) = P_1(u_1)P_2(u_2)$ holds in this ensemble of codes \cite{moradi2022tree}. 
This is corroborated by the experimental results shown in the next section, which says that the computational complexity of sequential decoding is unrelated to the CC.

Similar to the polar code, the PAC code with sequential decoding is a capacity-achieving code with low complexity. 
To see this, impose an upper constraint $B_N$ on the number of visits of sequential decoding during a decoding session in order to investigate the effect of the cutoff rate and channel polarization on decoding PAC codes.
Take, for example, the assumption that decoding will be stopped if the total number of visits exceeds $B_N$. 
Similarly, presume that $B_{N/2}$ is the upper constraint on the number of visits to the first half of the decoding tree levels and that decoding will be ended if the number of visits exceeds this upper bound. 
Continuing this way, the upper bound to decode the first bit is $B_1$.
By extreme limits, if $2B_i = B_{i+1}$ and $B_1 = 1$, the decoding is like the SC decoding.
Hence, this proves that the PAC code with infinite block length and a similar decoding complexity as SC decoding can achieve the channel capacity.

\begin{figure}[ht] 
\centering
	\includegraphics [width = \columnwidth]{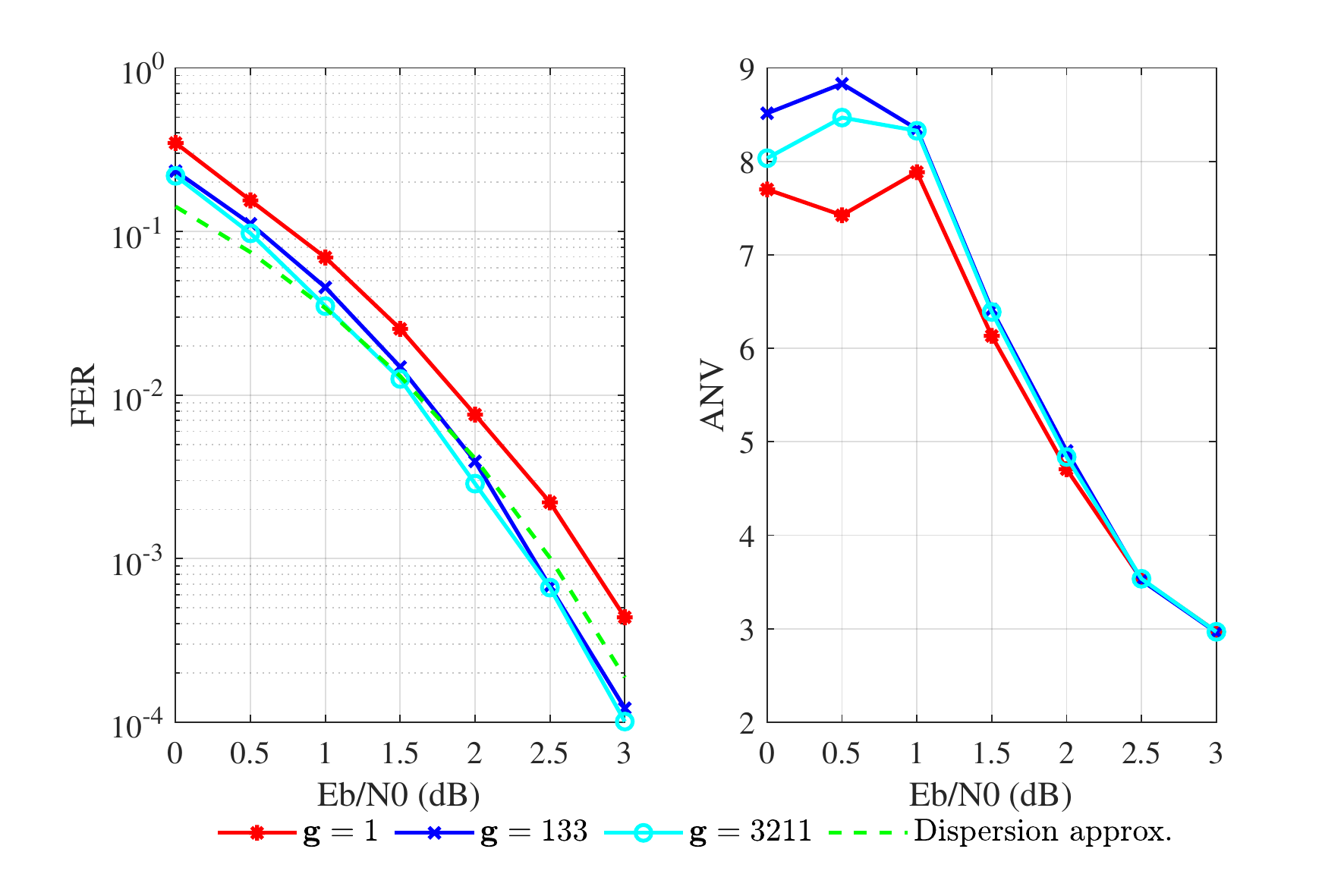}
	\caption{FER performance comparison of the PAC$(128, 29)$ codes with different connection polynomials.} 
	\label{fig: FER_ANV_K29N128}
\end{figure}

\begin{figure}[ht] 
\centering
	\includegraphics [width = \columnwidth]{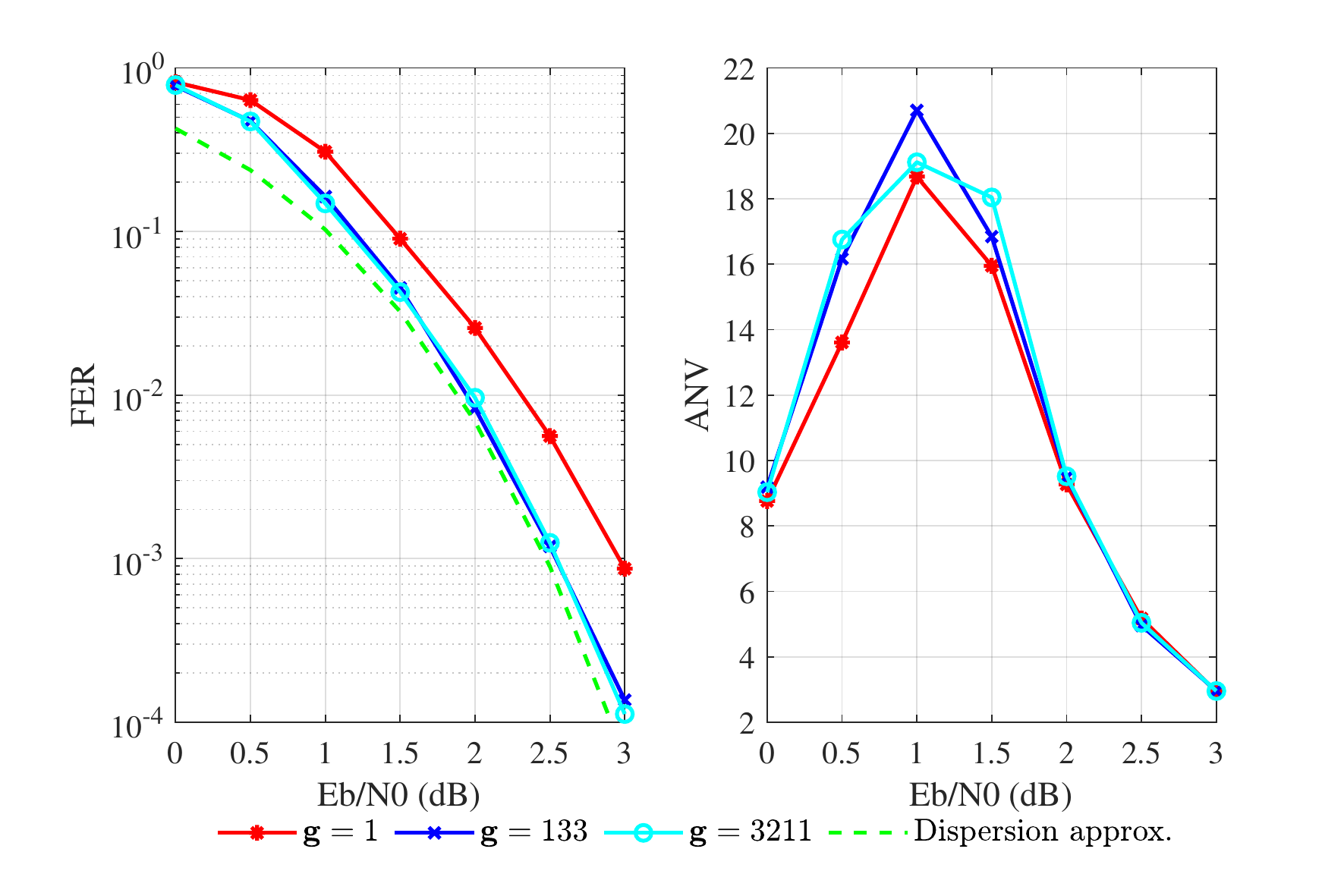}
	\caption{FER performance comparison of the PAC$(128, 64)$ codes with different connection polynomials.} 
	\label{fig: FER_ANV_K64N128}
\end{figure}

\begin{figure}[ht] 
\centering
	\includegraphics [width = \columnwidth]{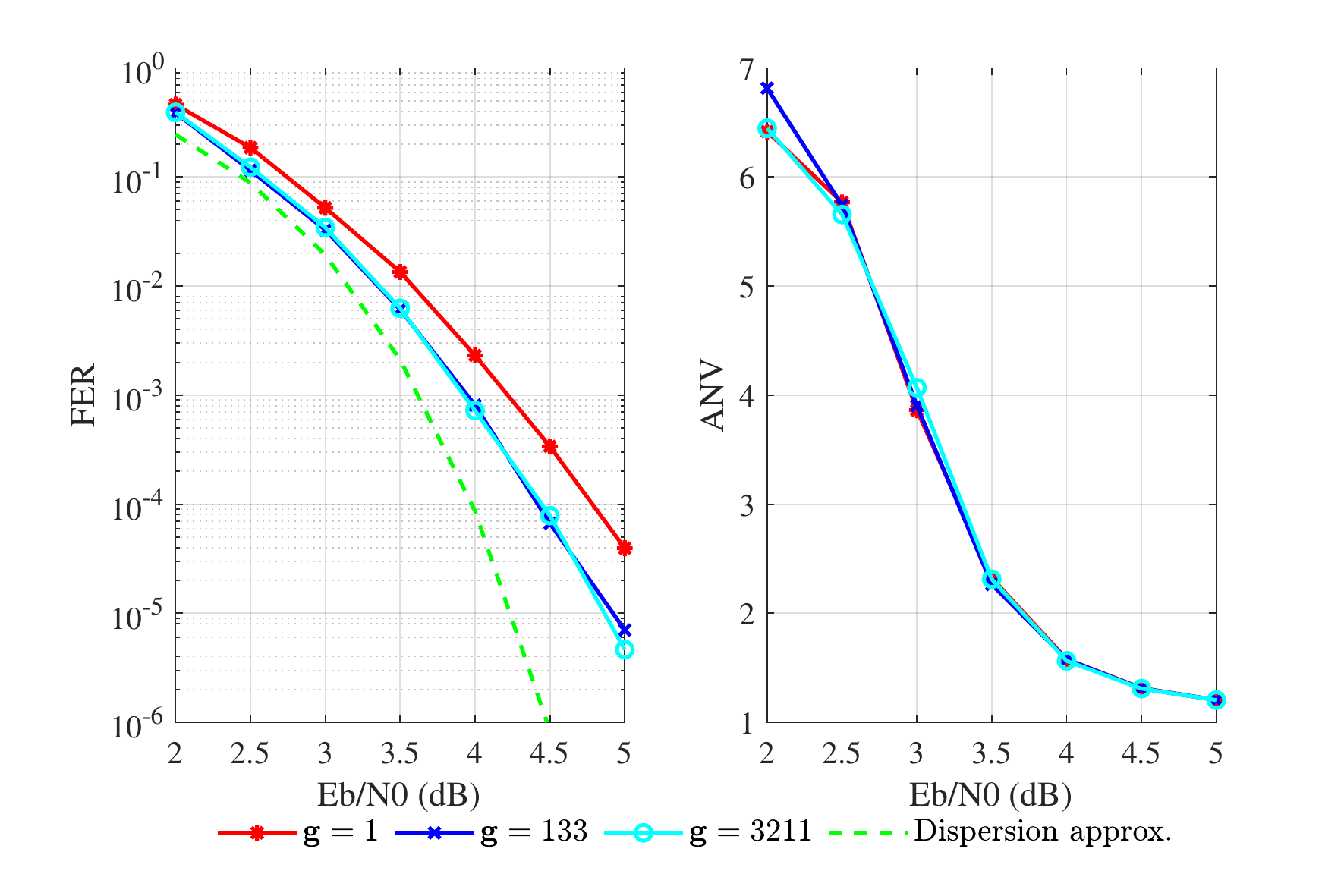}
	\caption{ FER performance comparison of the PAC$(128, 99)$ codes with different connection polynomials.} 
	\label{fig: FER_ANV_K99N128}
\end{figure}

\section{Simulation results} \label{sec: simulation}
The binary-input additive white Gaussian noise (BI-AWGN) channel with binary phase-shift keying (BPSK) modulation is considered in our simulations. 
We also compare our results with the frame error rate (FER) of the dispersion approximation \cite{polyanskiy2010channel}. 
The connection polynomials are presented in the octal form.
Fig. \ref{fig: FER_ANV_K29N128} provides a comparison between the FER performance of the PAC$(128, 29)$ codes ($\mathbf{g} = 133$ and $\mathbf{g} = 3211$) using RM rate-profile construction and the RM$(7, 2)$ code ($\mathbf{g} = 1$).
The RM code may be assumed to be a special instance of the PAC code with an identical convolutional encoder. 
This figure demonstrates that for all practical SNR levels, the ANV of the plots is almost the same, while there is a coding gain of around $0.5$~dB when employing a PAC code as opposed to an RM code with the same block length and code rate. 
Similar to Fig. \ref{fig: FER_ANV_K29N128}, Fig \ref{fig: FER_ANV_K64N128} and Fig. \ref{fig: FER_ANV_K99N128} compare the performance of the PAC codes with the corresponding RM codes at different code rates. 
According to these results, the PAC codes offer superior error-correction performance compared to the RM codes, even though their ANV values are almost similar.

\begin{figure}[ht] 
\centering
	\includegraphics [width = \columnwidth]{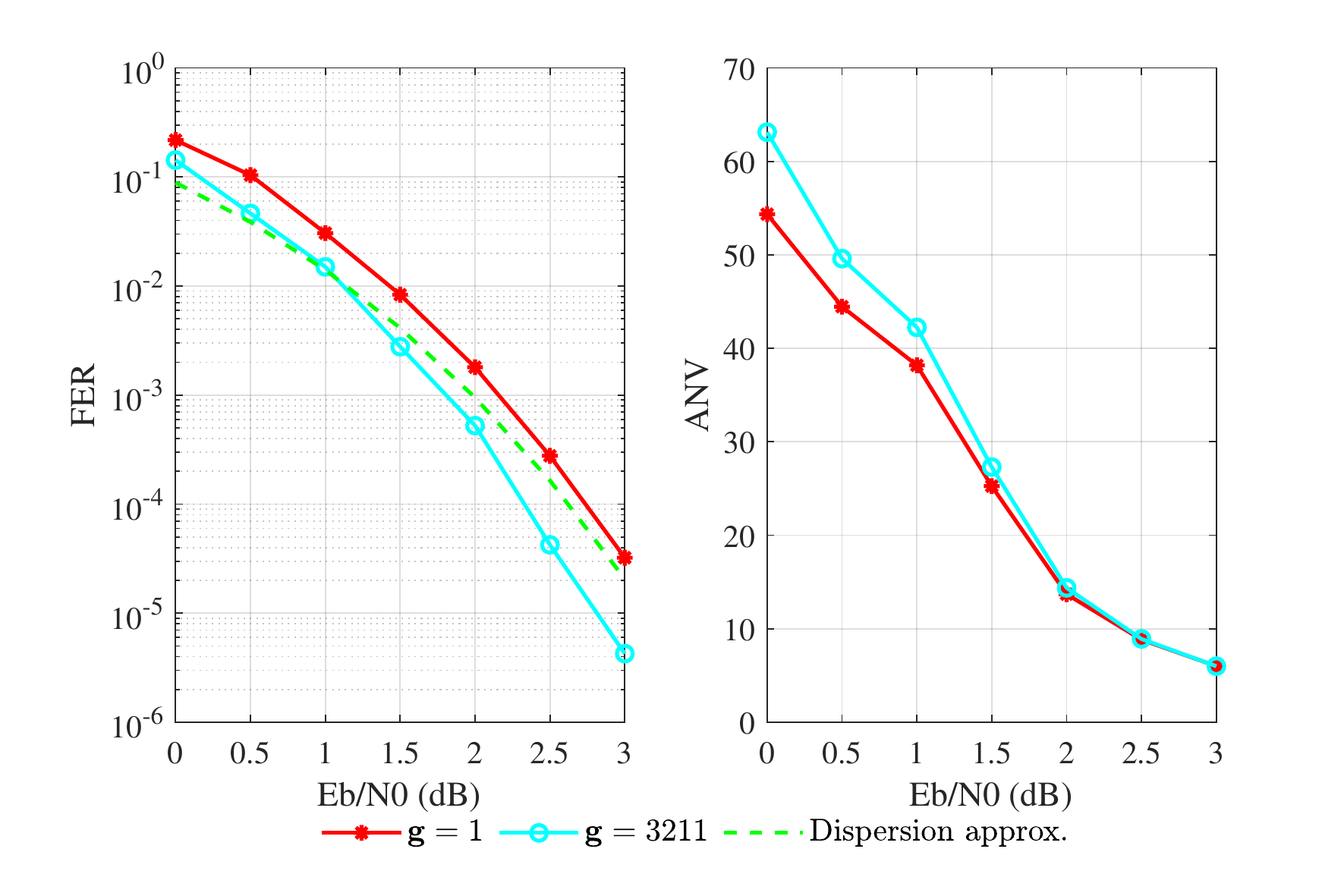}
	\caption{Performance comparison of the PAC$(256, 37)$ codes.} 
	\label{fig: FER_ANV_K37N256}
\end{figure}

\begin{figure}[ht] 
\centering
	\includegraphics [width = \columnwidth]{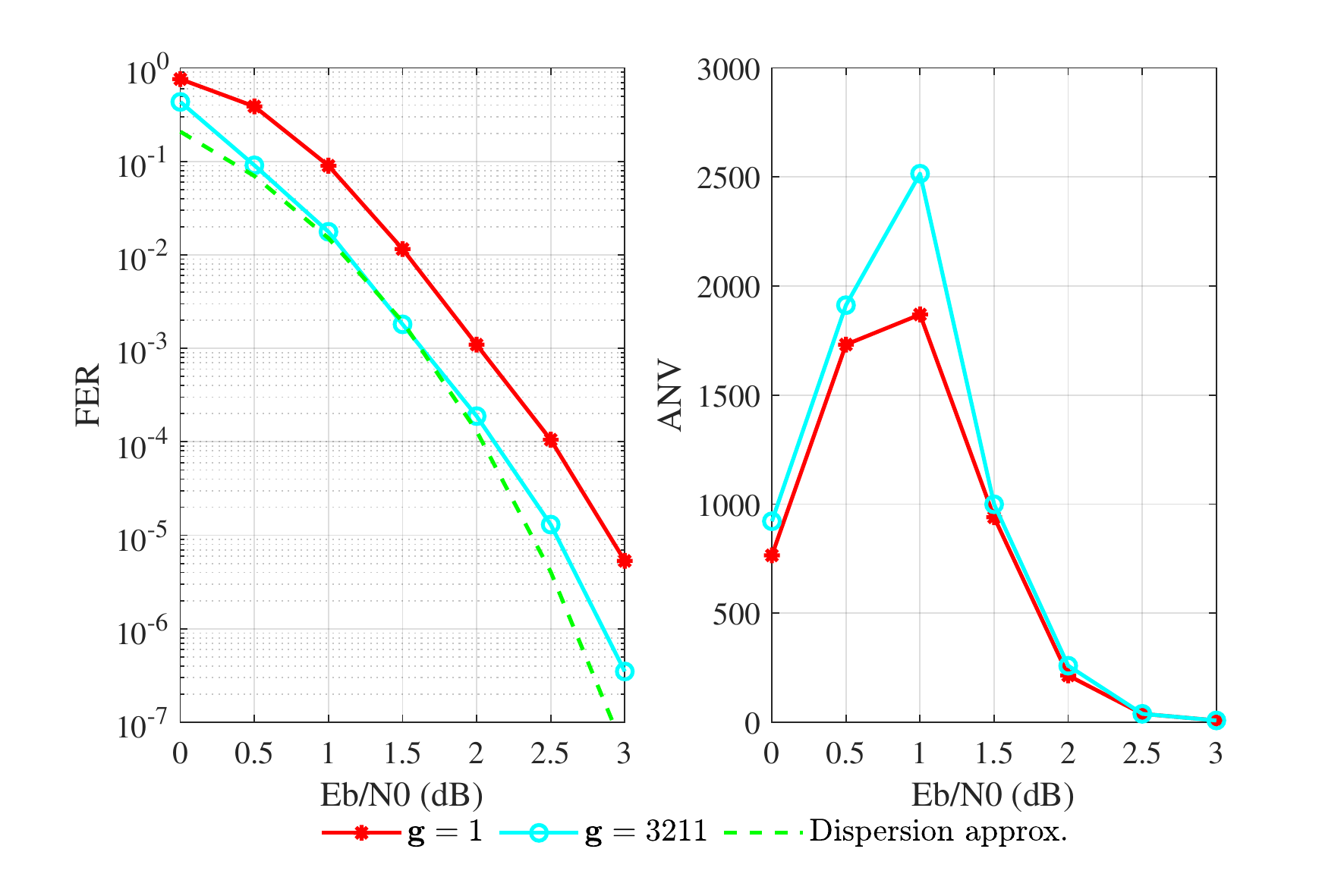}
	\caption{Performance comparison of the PAC$(256, 93)$ codes.} 
	\label{fig: FER_ANV_K93N256}
\end{figure}

\begin{figure}[ht] 
\centering
	\includegraphics [width = \columnwidth]{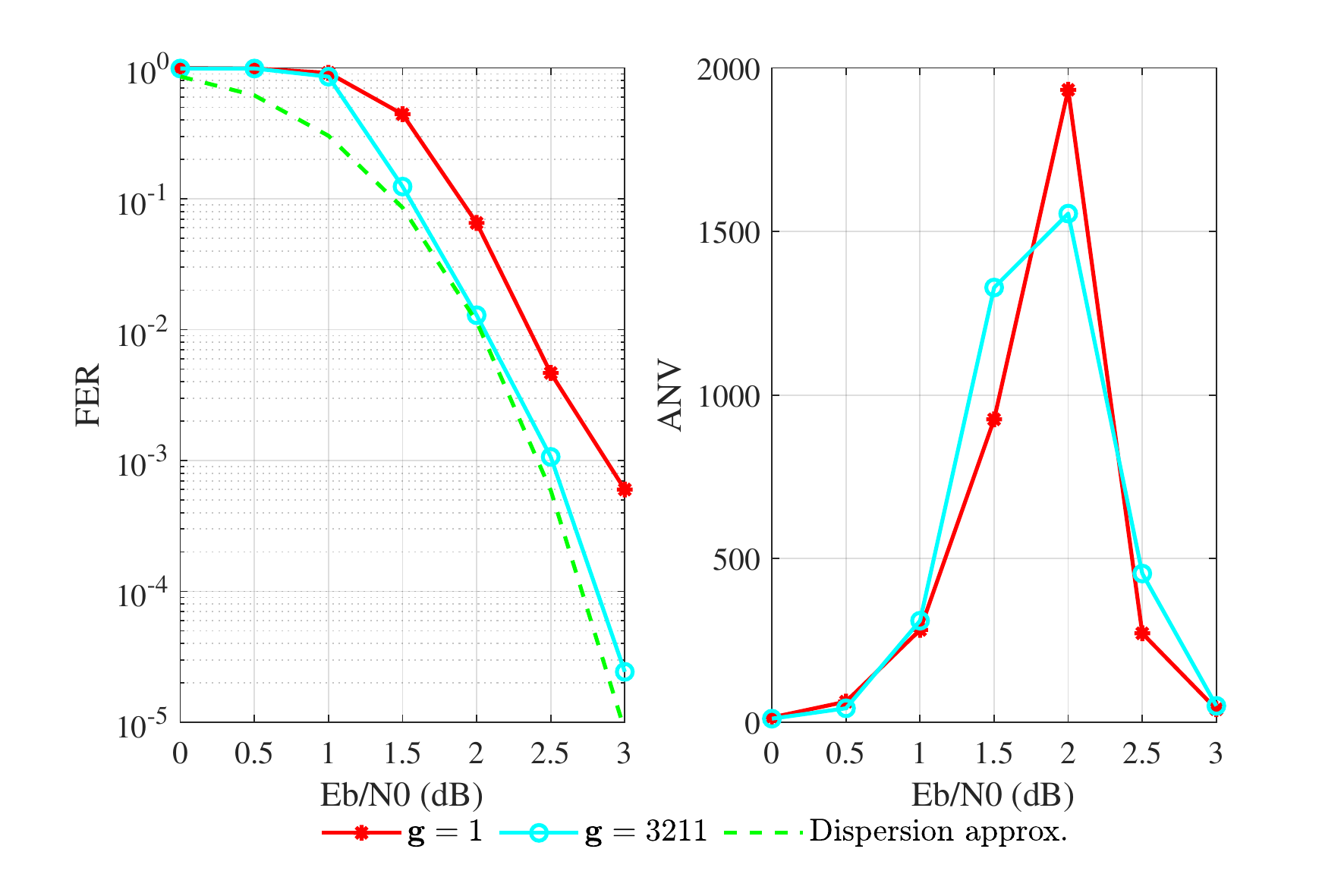}
	\caption{Performance comparison of the PAC$(256, 163)$ codes.} 
	\label{fig: FER_ANV_K163N256}
\end{figure}

Next, we study the performance of the $N=256$-length PAC codes ($\mathbf{g}=3211$) employing the RM rate profile construction and the RM codes ($\mathbf{g}=1$) in Fig. \ref{fig: FER_ANV_K37N256}, Fig. \ref{fig: FER_ANV_K93N256}, and Fig. \ref{fig: FER_ANV_K163N256}.
Similar to the $N=128$ cases, in $N=256$ the PAC codes have a coding gain of about $0.5$~dB over the RM codes, with almost the same ANV values. 
According to these figures, PAC codes using RM code construction can achieve the theoretical bounds for $N=256$ and $N=128$. 

As shown by the ANV figures, particularly for $N=256$, the computational complexity of a PAC code with the RM rate-profile construction can be extremely high. 
Following this, we will attempt to tame the RM rate-profile construction by benefiting from the guessing technique. 
We explain our taming method by providing a detailed example.

\begin{figure*}[ht] 
\centering
	\includegraphics [width = .8\textwidth]{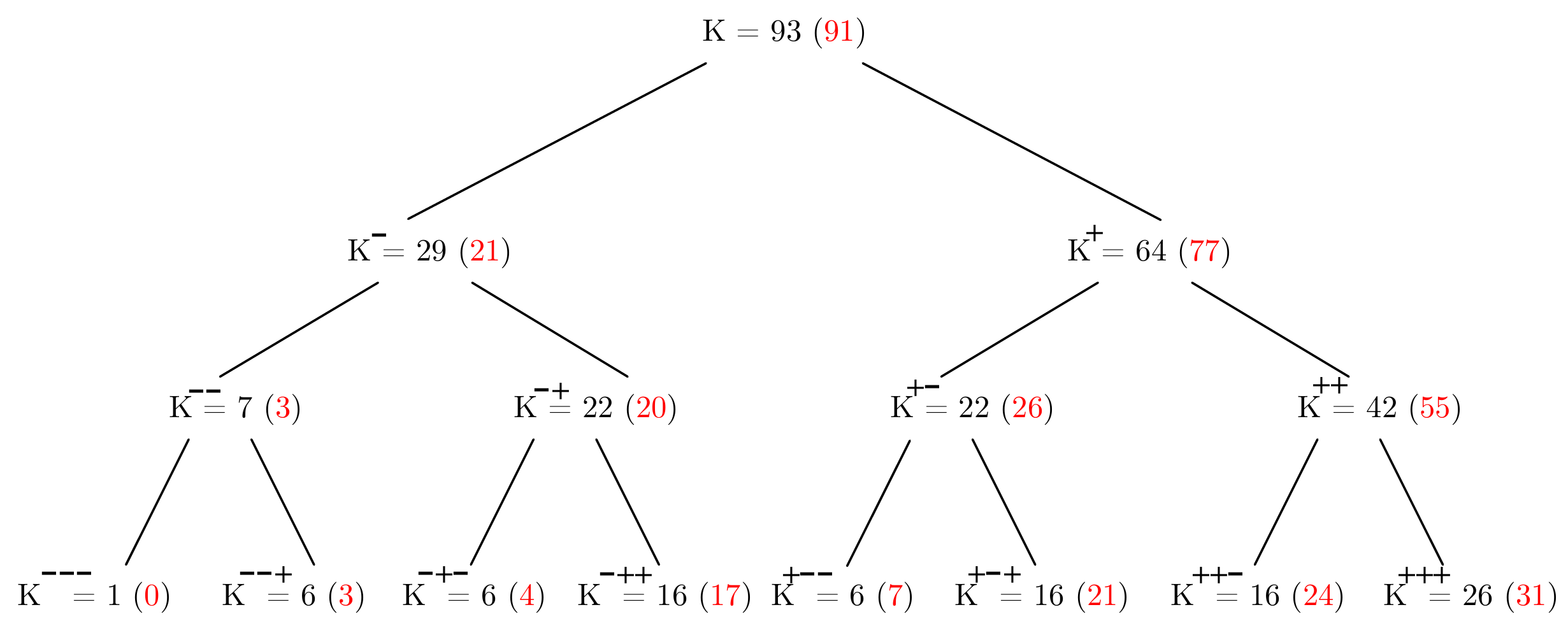}
	\caption{Polarization tree for the length of the information bits for the PAC$(256, 93)$ codes.} 
	\label{fig: PolarizationTree}
\end{figure*}

\begin{figure}[ht] 
\centering
	\includegraphics [width = \columnwidth]{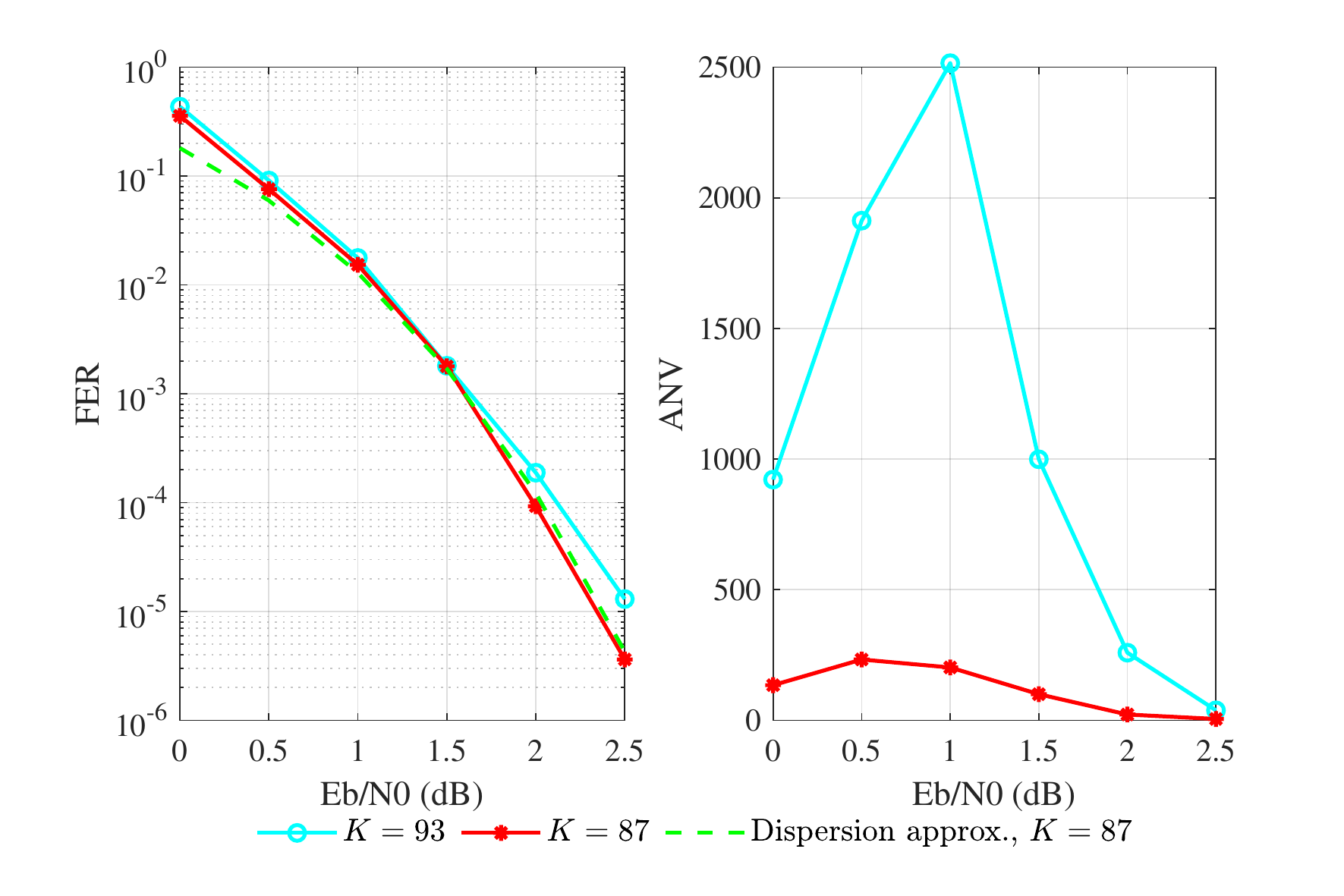}
	\caption{Performance comparison of the PAC codes of length $N = 256$ with different data lengths.} 
	\label{fig: Tame_K93N256}
\end{figure}

As an example, let us consider a PAC$(256, 93)$ code at $E_b/N_0 = 2$~dB.
In this case, $R_0(W) = 0.3564$.
For the sake of approximation, assume $\epsilon = 0.1$. 
Based on \eqref{eq: N_guess_LB}, $K$ needs to be about smaller than $\lfloor NR_0(W) + \epsilon\rfloor = 91$. 
To meet the condition \eqref{eq: N_guess_LB}, we may remove $2$ bit indices from the RM rate profile of the PAC$(256, 93)$ code.

In the PAC$(256, 93)$ code, $K^- = 29$ and $K^+ = 64$.
Also, after one step of polarization, $\lfloor \frac{N}{2}R_0(W^-) + \epsilon\rfloor = 21$ and $\lfloor \frac{N}{2}R_0(W^+) + \epsilon\rfloor = 77$.
In order to satisfy condition \eqref{eq: badN/2} in the RM rate-profile construction, $8$ bit indices contributing to the first half of the rate profile must be frozen. 
In our simulations, we do freezing from the indexes with the smallest positions.
Fig. \ref{fig: PolarizationTree} depicts the polarization tree of the information bit length. 
The red values in the parentheses represent the largest amount of information bits allowed to fulfill the lower bound equations using the guessing technique.
This tree displays three levels of polarization.
For instance, if we examine the third level of polarization, the first node of the tree indicates that the first $32$ bits all must be frozen bits. Note that one of the first $32$ bits in the RM code construction is an information bit, and we should freeze this bit.
The second $32$ bits of the RM construction also include $6$ information bits, while the guessing technique requires us to have just $3$ of these bits be information bits and we should freeze $3$ of them. 
In addition, $2$ more information bits must be frozen from the third $32$ bits. 
This means that totally by freezing $6$ bits, the guessing technique's polarized computational cutoff rate limit will be met at the third polarization level. 

For this PAC$(256, 93)$ code, freezing the $\{32, 48, 56, 60, 80, 88 \}$ bit locations would satisfy the third step polarization limit of the computational cutoff rate and results in a PAC$(256, 87)$ code.
Fig. \ref{fig: Tame_K93N256} compares the performance of the PAC$(256, 93)$ code and the obtained PAC$(256, 87)$ code.
The ANV results show that our proposed technique results in a significant complexity reduction.

Note that on the third level of this tree, for instance, the seventh node can contain $\textcolor{red}{24}$ bits of information, but the RM code construction consists of just $16$ bits.
This will be used to generalize the tamed RM code construction to an arbitrary code rate in the following section.

\begin{figure}[ht] 
\centering
	\includegraphics [width = \columnwidth]{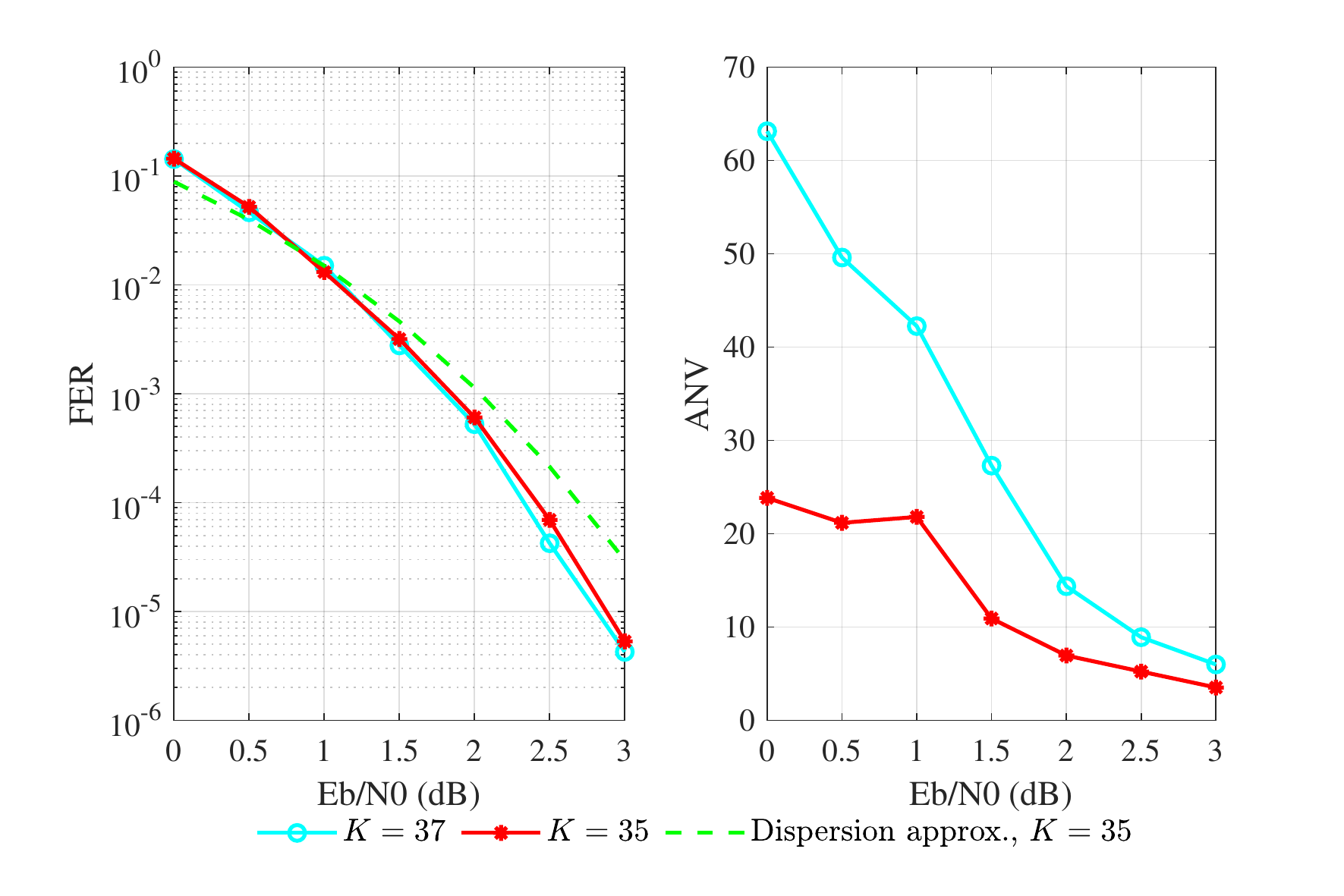}
	\caption{Performance comparison of the PAC codes of length $N = 256$ with different data lengths.} 
	\label{fig: Tame_K37N256}
\end{figure}

\begin{figure}[ht] 
\centering
	\includegraphics [width = \columnwidth]{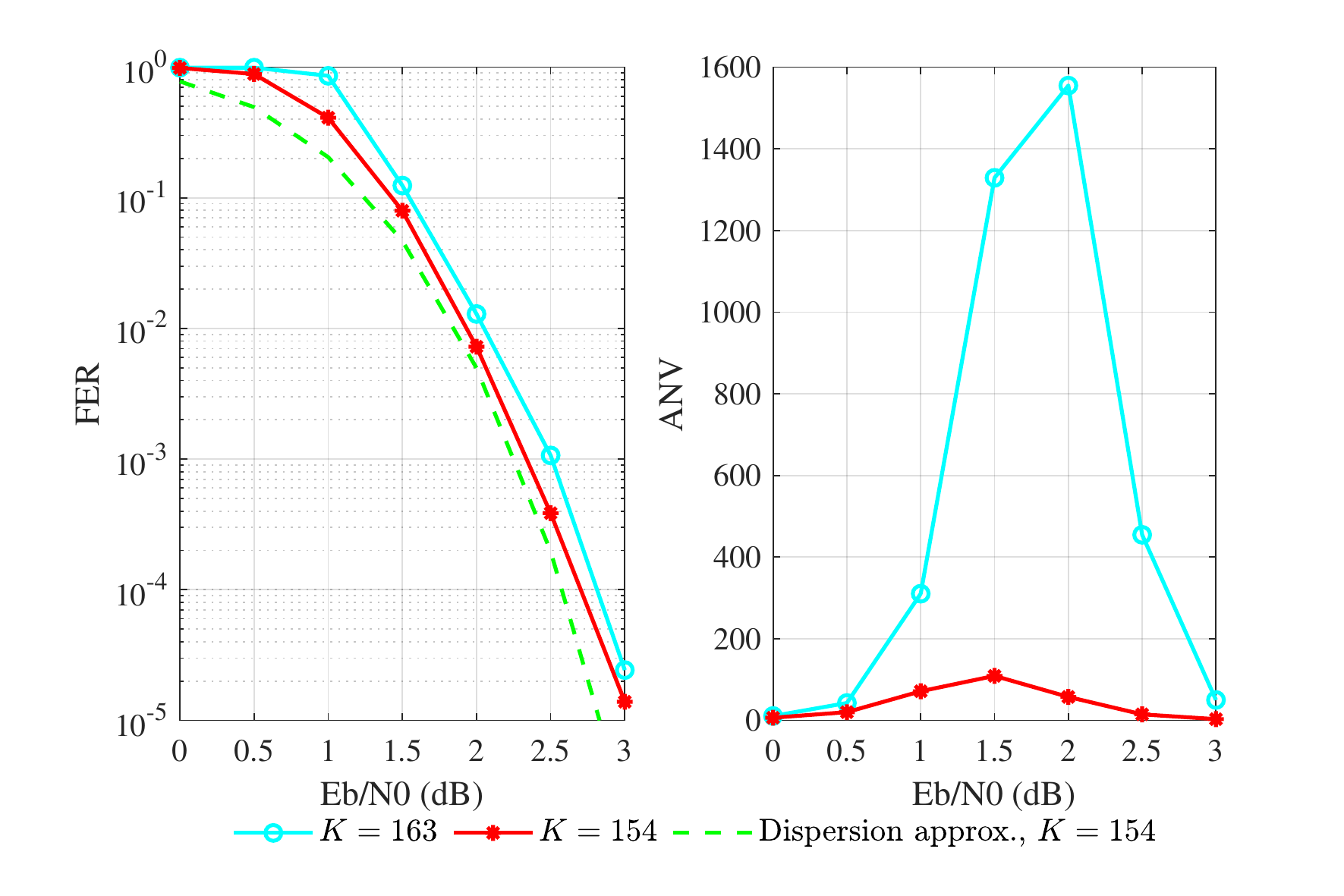}
	\caption{Performance comparison of the PAC codes of length $N = 256$ with different data lengths.} 
	\label{fig: Tame_K163N256}
\end{figure}

For a low rate PAC$(256, 37)$ code, freezing the bit locations that satisfy the third step polarization limit of the computational cutoff rate at a $E_b/N_0 = 2.5$~dB results in a PAC$(256, 35)$ code.
Fig. \ref{fig: Tame_K37N256} compares the performance of the PAC$(256, 37)$ code and the obtained PAC$(256, 35)$ code.
Similarly, Fig. \ref{fig: Tame_K163N256} obtains a PAC$(256, 154)$ code from the high rate PAC$(256, 163)$ code at $E_b/N_0 = 2$~dB.
The ANV results in both plots indicate that our suggested approach leads to a substantial reduction of complexity.

\begin{figure}[ht] 
\centering
	\includegraphics [width = \columnwidth]{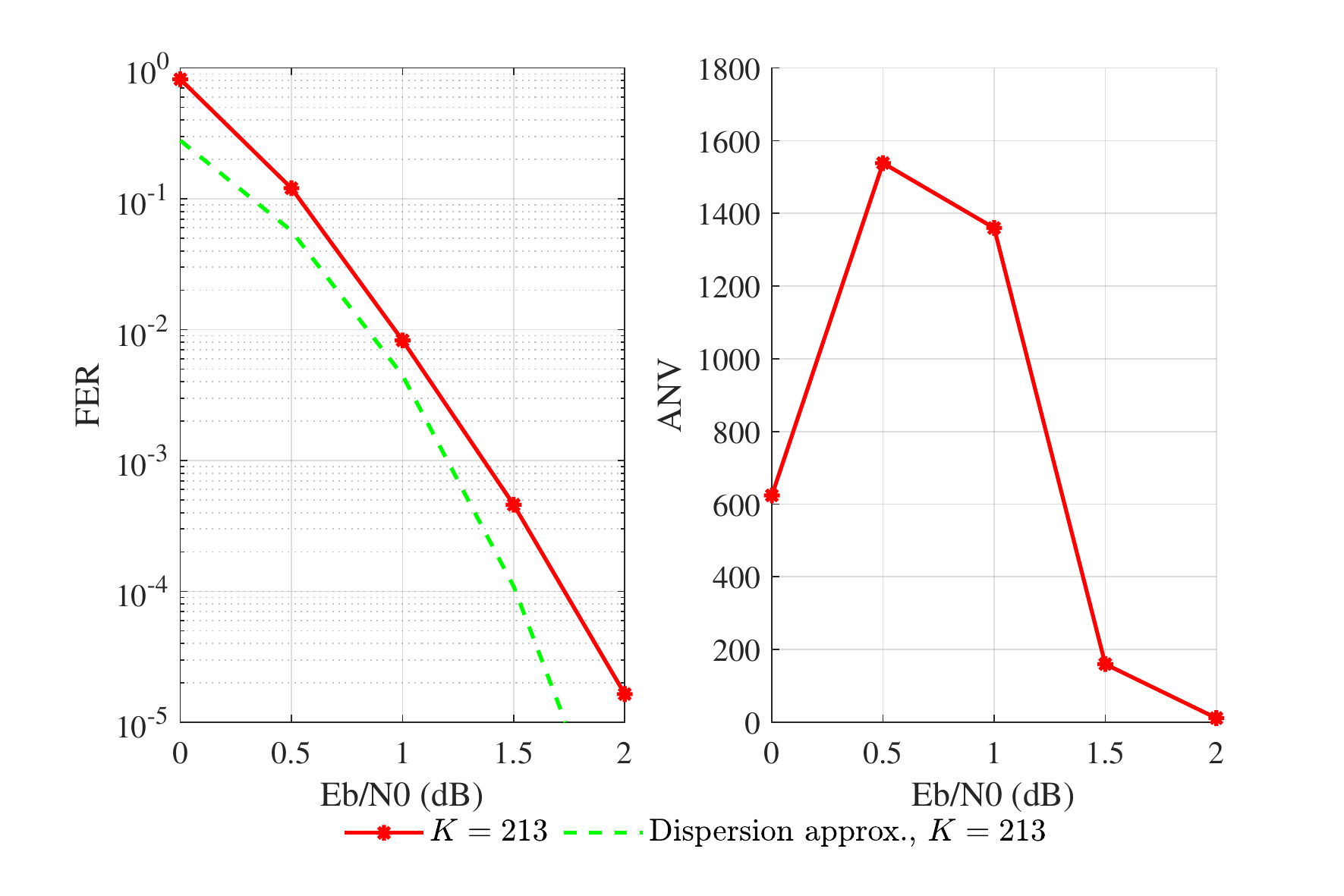}
	\caption{Performance comparison of the PAC$(512,213)$ code and the $(512,256)$ polar code.} 
	\label{fig: Tame_K256N512}
\end{figure}

Sequential decoding of moderate-length polar or PAC codes employing RM code construction has an extremely high computational complexity, and polar code construction \cite{arikan2009channel} is typically employed as the code construction. 
For this reason, we will only present the performance of the tamed RM code construction for the $N=512$ PAC codes in this part.
By freezing the bit positions of the RM$(512, 256)$ code that fulfills the fourth step polarization limit of the computational cutoff rate at $E_b/N_0 = 1$~dB, the PAC$(256, 213)$ code can be obtained. 
Fig. \ref{fig: Tame_K256N512} depicts the performance of the PAC$(256, 213)$ code together with its corresponding dispersion approximation.
As this figure shows, the FER performance of the PAC code is very close to the dispersion approximation plot.

\begin{figure}[ht] 
\centering
	\includegraphics [width = \columnwidth]{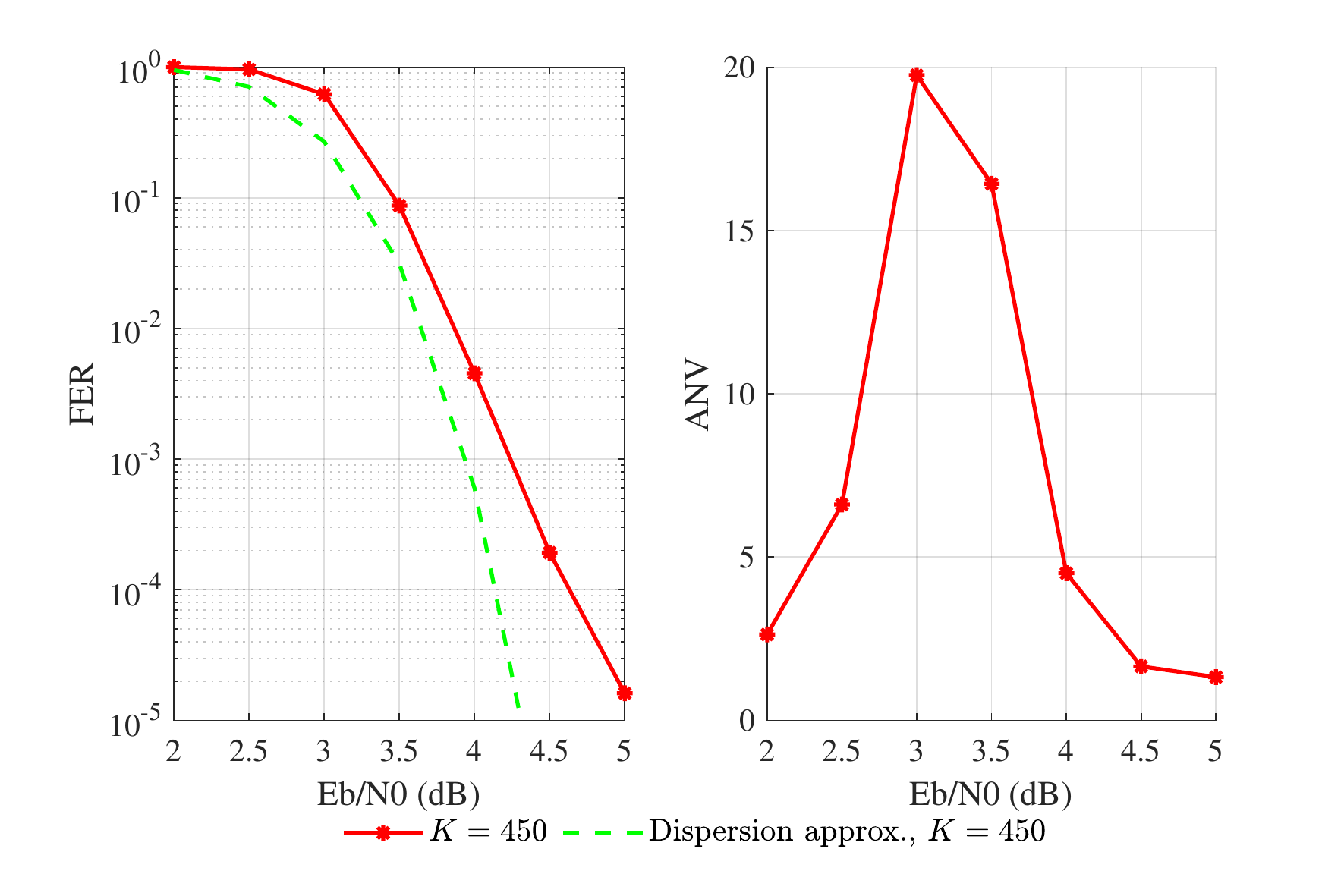}
	\caption{Performance of the PAC$(512,450)$ code.} 
	\label{fig: Tame_K466N512}
\end{figure}

\begin{figure}[ht] 
\centering
	\includegraphics [width = \columnwidth]{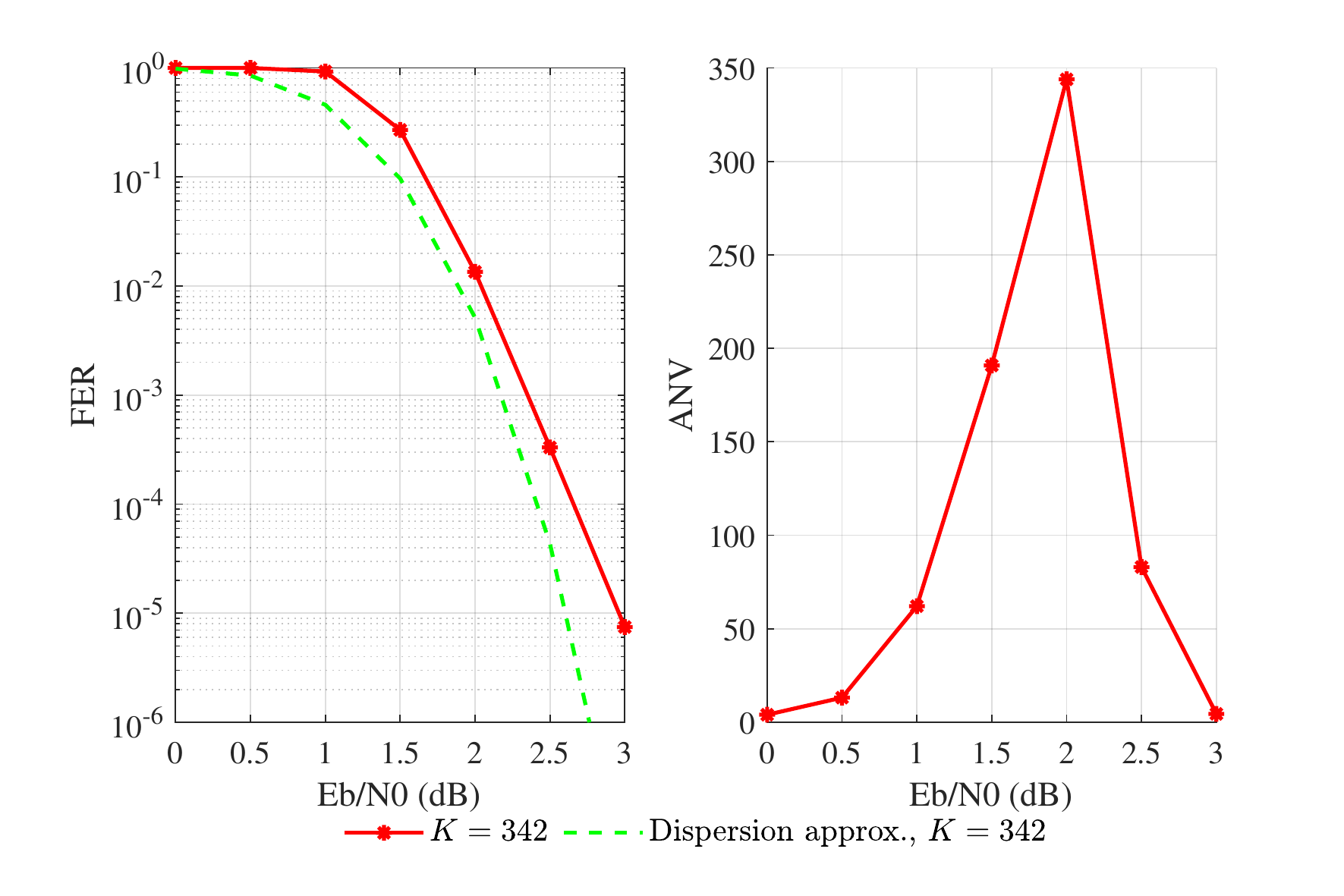}
	\caption{Performance of the PAC$(512,342)$ code.} 
	\label{fig: Tame_K382N512}
\end{figure}

For the RM$(512, 466)$ code, freezing the bit locations that satisfy the fourth step polarization limit of the computational cutoff rate at a $E_b/N_0 = 4$~dB results in a PAC$(512, 450)$ code.
Fig. \ref{fig: Tame_K466N512} plots the ANV and FER performance of this obtained code.
Similarly, Fig. \ref{fig: Tame_K382N512} obtains a PAC$(512, 342)$ code from the high rate RM$(512, 382)$ code at $E_b/N_0 = 2$~dB.
In both figures, the error-correction findings show that the PAC code constructed by our proposed tamed-RM rate profile construction is close to the theoretical bounds and has a low computational complexity at high $E_b/N_0$ values. 
Note that the taming of the RM rate profile construction at the lower SNR values leads in a better computational complexity at the price of a reduced PAC coding rate.


\subsection{Arbitrarily Code Rates}\label{sec: rate}
The rate choices for the RM code are limited. 
To obtain an arbitrary rate PAC code, we provide a heuristic method with two examples. 

To obtain a rate profile for the PAC$(512, 256)$ code, we use the RM$(512, 256)$ and RM$(512,382)$ codes. 
Note that the RM$(512, 256)$ code has a minimum distance of $32$ and the RM$(512, 382)$ code has a minimum distance of $16$. 
As noted in the previous section, employing all of the bit indices corresponding to the RM$(512, 256)$ code will result in sequential decoding with high computational complexity. 
By discarding the bit indices based on the cutoff rate polarization requirements set by the level of $5$ at $1.5~$dB SNR value, a PAC$(512, 230)$ code can be generated. 
Similarly, by eliminating the bit indices corresponding to RM$(512, 382)$ code based on the level of $5$ at $0.5~$dB SNR value, a PAC$(512, 290)$ code can be obtained.

To get a rate profile for a PAC$(512, 256)$ code, we choose all of the bit indices from the PAC$(512, 230)$ code and the remaining $26$ bit indices from the PAC$(512, 290)$ code whose rows have weights of $16$. 
Then, choose the lowest bit indices of rows with weights of $16$ that fulfill the cutoff rate criterion of PAC$(512, 230)$ at level $5$. 
The newly added bit indexes for this example are
[211, 227, 229, 241, 307, 309, 326, 327, 338, 339, 341, 345, 354, 355, 357, 361, 369, 388, 390, 391, 394, 402, 403, 405, 409, 418].

\begin{figure}[ht] 
\centering
	\includegraphics [width = \columnwidth]{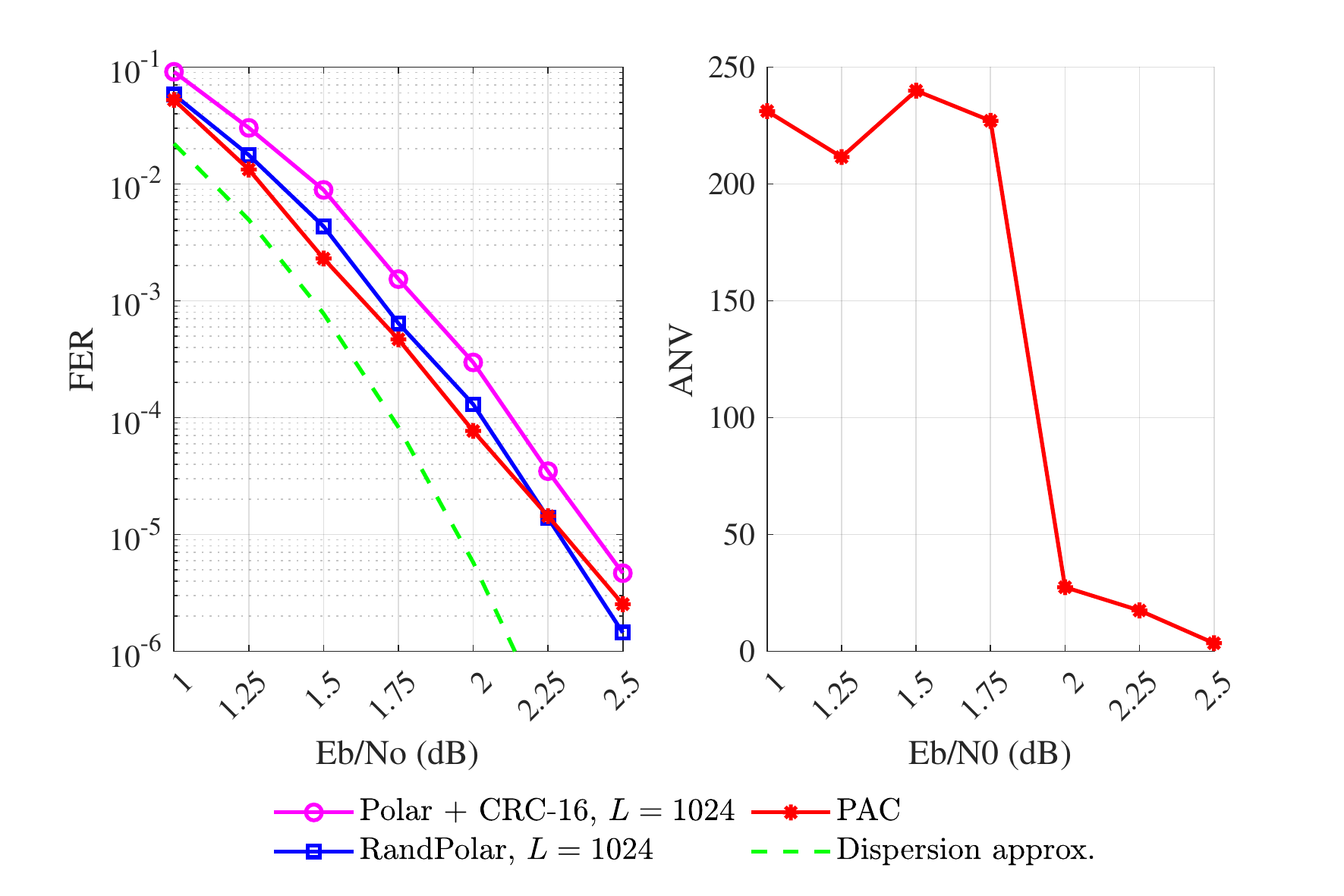}
	\caption{Performance comparison of the PAC$(512, 256)$ code and the $(512, 256)$ polar codes.} 
	\label{fig: Tame_K256N512_RM_Guessing}
\end{figure}

The ANV and FER performance of this resulting code is plotted in Fig. \ref{fig: Tame_K256N512_RM_Guessing}. 
The performance of the CRC-aided list decoding polar code with a CRC length of $16$ and a list size of $1024$ is also plotted in this figure.
For comparison, the performance of the list decoding of the polar code of \cite{cocskun2022information} constructed for the list size of 1024 and designed based on the polarization weight, in which each frozen bit is set to a random linear combination of the previous information bits, is also plotted in this figure. 
Note that at $2.5~$dB, our construction technique's ANV value is less than $3.48$, which is much less than $1024$.

Similarly, we employ the RM$(256, 93)$ and RM$(256, 163)$ codes to construct a rate profile for the PAC$(256, 128)$ code. 
Note that there is no RM code of rate $0.5$ for a code length of $N = 256$. 
The RM$(256, 93)$ code has a minimum distance of $32$ and the RM$(512, 163)$ code has a minimum distance of $16$. 
Starting with the RM$(256, 93)$ code, we can obtain the PAC$(256, 92)$ code by removing the bit indices based on the cutoff rate polarization requirements set at the level of $4$ at $3.5~$dB SNR value. 
Similarly, PAC$(256, 150)$ code can be obtained by removing the bit indices corresponding to RM$(256, 163)$ code at the level of $4$ at $2~$dB SNR value. 
To construct the rate profile for a PAC$(256, 128)$ code, we choose all of the bit indices from the PAC$(256, 92)$ code and the remaining $36$ bit indices from the PAC$(256, 150)$ code whose rows have weights of $16$. 
Out of those bit indices with weight $16$, we select the lowest bit indices that fulfill the polarization requirement at level $4$ imposed by the PAC$(256, 92)$ code.
The newly introduced bit indices to the PAC$(256, 92)$ code for this example are 
[58, 78, 84, 86, 87, 100, 102, 103, 106, 114, 115, 117, 121, 136, 140, 148, 150, 151, 154, 155, 164, 166, 167, 170, 171, 173, 178, 179, 181, 185, 196, 198, 199, 202, 203, 205].

\begin{figure}[ht] 
\centering
	\includegraphics [width = \columnwidth]{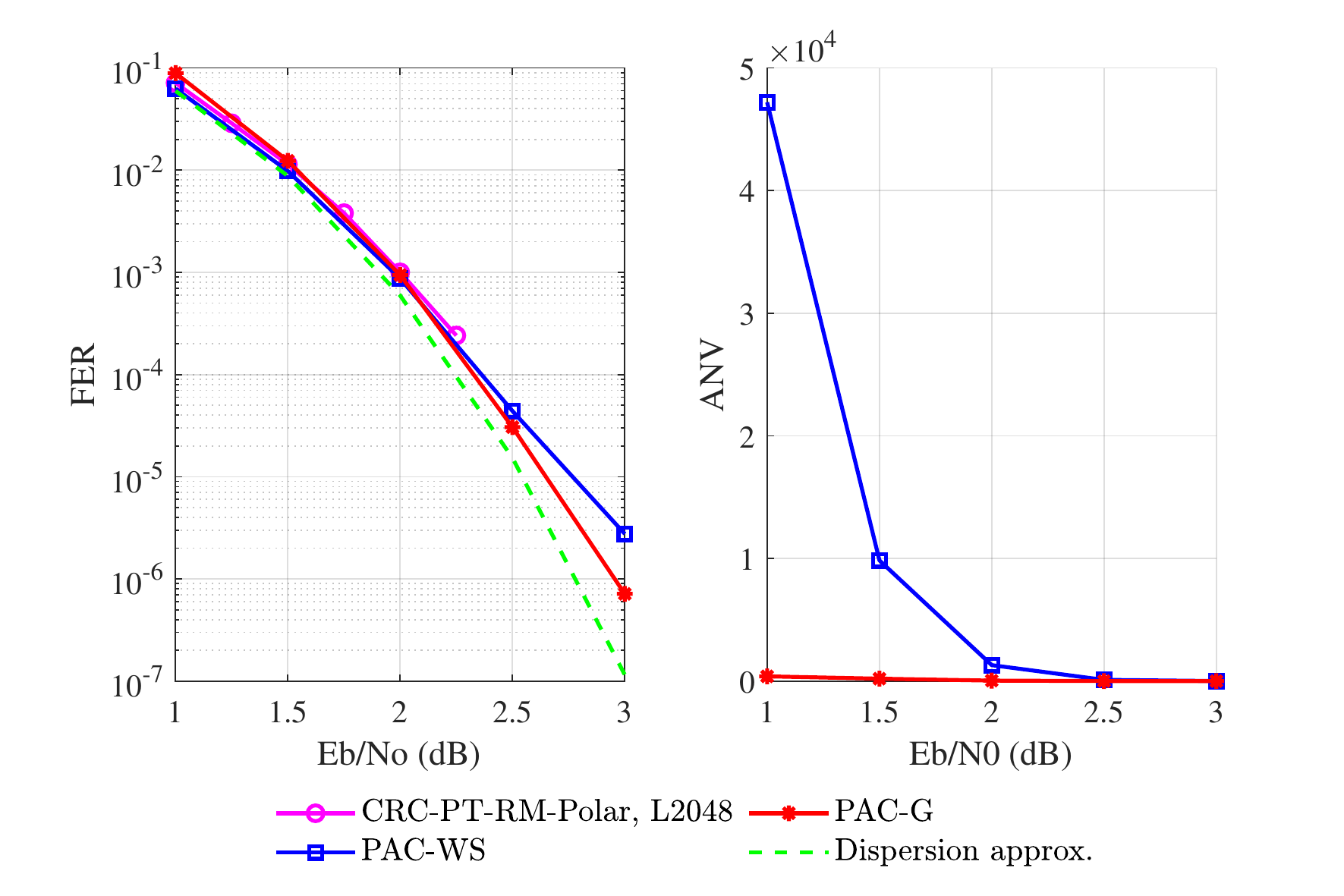}
	\caption{Performance comparison of the PAC$(256, 128)$ codes and the $(256, 128)$ polar code.} 
	\label{fig: Tame_K128N256_RM_Guessing}
\end{figure}

The ANV and FER performance of this code (PAC-G) is plotted in Fig. \ref{fig: Tame_K128N256_RM_Guessing}. 
The performance of the CRC-aided list decoding of the pre-transformed RM-polar code of \cite{li2021performance} with a CRC length of $6$ and a list size of $2048$ is plotted in this figure. 
The performance of the Fano decoding of the PAC code of \cite{liu2022weighted}, whose rate profile is constructed using a weighted sum (PAC-WS) is also plotted in this figure for comparison. 
The ANV values of our construction approach at $1~$dB SNR is about $391$ and at $3~$dB it is nearly $2.82$, while the ANV value of PAC-WS code is nearly $9.8$ at $3~$dB.

\section{Conclusion}\label{sec: conclusion}
In this paper, we investigated the performance of the sequential decoding of the PAC codes constructed by the RM rate profile.
We proved the polarization of the computational cutoff rate in the sequential decoding of PAC codes.
We also proposed a technique for taming the computational complexity of sequential decoding based on the polarization of the computational cutoff rate. 
Simulation results demonstrate that our rate profile design enables the FER performance of the PAC code to meet the theoretical bounds at moderate code block lengths with much less computational complexity than when using the RM rate profile. 
For an RM$(r,m)$ code only a limited number of code rates are available, while we have generalized our approach to arbitrary code rates.

\section*{Acknowledgment}
I would like to thank Professor Erdal Ar{\i}kan for his guidance during this work.

\ifCLASSOPTIONcaptionsoff
  \newpage
\fi

\bibliographystyle{IEEEtran}
\bibliography{bibliography}

\end{document}